\documentclass[11pt]{article}

\usepackage{slashed}
\usepackage{color}
\usepackage{graphicx}  % standard LaTeX graphics tool
\usepackage[all]{xypic}
\usepackage{psfrag}
\usepackage{tocbibind} %adds contents and bibliography to the table of contents
\usepackage{epsf}
\usepackage{epic}
\usepackage{eepic}
\usepackage{epsfig}
\usepackage{wrapfig}
\usepackage{a4}
\usepackage{amssymb,latexsym}
\usepackage[mathscr]{eucal}
\usepackage{cite}
\usepackage[active]{srcltx}
\usepackage{url}
\usepackage{fancybox}
\usepackage{ntheorem}
\usepackage{mhequ}
\usepackage[g]{esvect}
\usepackage{epstopdf}

\usepackage[utf8]{inputenc}

\usepackage{fourier}
\usepackage{braket}
\usepackage{epstopdf}
\usepackage{mathrsfs}
\usepackage{stmaryrd}
\usepackage{upgreek}
\usepackage{floatrow}
\usepackage{xfrac}
\usepackage{subfigure}

\usepackage{tikz}

\usepackage{cancel}
\usepackage{float}
\usepackage{enumitem}
\usepackage{bbm}
\usepackage{caption}

\usepackage[utf8]{inputenc}

\newcommand{\barvarphi}{\bar \varphi}
\newcommand{\phinotone}{\phi}
\newcommand{\alphanotone}{\alpha}

\newcommand{\RPtwo}{{\mathbbm{R}\mathbbm{P}^2}}
\newcommand{\PRtwo}{\RPtwo}

\newcommand{\lnn}{\sigma \frac{\hat n_1}{\hat n_2}}

\newcommand{\Mastro}{M_{\textrm{astro}}}
\newcommand{\Xilor}{\Xi_{{{\mathrm{Lor}}}}}
\newcommand{\Deltalor}{\Delta_{{{\mathrm{Lor}}}}}

\newcommand{\ephys}{{%\color{red}
     q}}%p_{\textrm{eff,Lor}}}}
\newcommand{\ephysphys}{{%\color{red}
    |\ephys_{\phys}|}}%p_{\textrm{eff,Lor}}}}

\newcommand{\esquare}{p^2_{\textrm{eff}}}
\newcommand{\noQ}{{\esquare}}

\newcommand{\nmax}{n_{\max{}}}

\newcommand{\phys}{\textrm{phys}}
\newcommand{\almostkappa}{\lambda}

\newcommand{\eqref}[1]{\eq{#1}}

\newcommand{\hs}{\cH_{\mbox{\scriptsize sing}}}

\newcommand{\beadl}[1]{\begin{deqarr}\label{#1}}
\newcommand{\eeadl}[1]{\arrlabel{#1}\end{deqarr}}%
\def\nz{\ifmmode {I\hskip -3pt N} \else {\hbox {$I\hskip -3pt N$}}\fi}
\def\zz{\ifmmode {Z\hskip -4.8pt Z} \else
       {\hbox {$Z\hskip -4.8pt Z$}}\fi}
\def\qz{\ifmmode {Q\hskip -5.0pt\vrule height6.0pt depth 0pt
       \hskip 6pt} \else {\hbox
       {$Q\hskip -5.0pt\vrule height6.0pt depth 0pt\hskip 6pt$}}\fi}
\def\rz{\ifmmode {I\hskip -3pt R} \else {\hbox {$I\hskip -3pt R$}}\fi}
\def\cz{\ifmmode {C\hskip -4.8pt\vrule height5.8pt\hskip 6.3pt} \else
       {\hbox {$C\hskip -4.8pt\vrule height5.8pt\hskip 6.3pt$}}\fi}
\def\au{{\setbox0=\hbox{\lower1.36775ex\hbox{''}\kern-.05em}\dp0=.36775ex\hs
kip0pt\box0}}
\def\ao{{}\kern-.10em\hbox{``}}

\newcommand\Gregbeq{\begin{eqnarray}}
\newcommand\Gregeeq{\end{eqnarray}}

\newcommand{\f}{\partial}

\def\cH{{\cal H}}

\def\h1{{\hat 1}}
\def\h2{{\hat 2}}

\def\3f{\frac{3}{2}}

\newcommand{\roscoff}[1]{}

{\catcode `\@=11 \global\let\AddToReset=\@addtoreset}
\AddToReset{figure}{section}

{\catcode `\@=11 \global\let\AddToReset=\@addtoreset}
\AddToReset{table}{section}

\DeclareFontFamily{OT1}{rsfs}{}
\DeclareFontShape{OT1}{rsfs}{m}{n}{ <-7> rsfs5 <7-10> rsfs7 <10-> rsfs10}{}
\DeclareMathAlphabet{\mycal}{OT1}{rsfs}{m}{n}

{\catcode `\@=11 \global\let\AddToReset=\@addtoreset}
\AddToReset{equation}{section}

\newcounter{mnotecount}[section]

\renewcommand{\themnotecount}{\thesection.\arabic{mnotecount}}

\newcommand{\oversetty}[2]{%
\mathop{#2}\limits^{\vbox to -.1ex{%
\kern -1.5ex\hbox{$\scriptstyle #1$}\vss}}}

\newcommand{\jlcax}[1]{}
\newcommand{\eean}{\nonumber\end{eqnarray}}

\newcommand{\kk}[1]{}%{\mnote{{\bf If we consider the KK case:} #1}}

\newcommand{\beq}{\begin{equation}}

\newcommand{\FS}       %{F_1} %
                  {F}
\newcommand{\HS} %{F_2}
       {H_{\mbox{\scriptsize volume}}}

{\ptc{this should be removed in the oberwolfach version}}%

\newcommand{\eel}[1]{\label{#1}\end{equation}}
\newcommand{\eeal}[1]{\label{#1}\end{eqnarray}}
\newcommand{\bed}{\begin{deqarr}}
\newcommand{\eed}{\end{deqarr}}
\newcommand{\bedl}[1]{\begin{deqarr}\label{#1}}
\newcommand{\eedl}[2]{\arrlabel{#1}\label{#2}\end{deqarr}}

\newcommand{\bel}[1]{\begin{equation}\label{#1}}
\newcommand{\bea}{\begin{eqnarray}}
\newcommand{\bean}{\begin{eqnarray}\nonumber}
\newcommand{\beal}[1]{\begin{eqnarray}\label{#1}}
\newcommand{\eea}{\end{eqnarray}}

\newcommand{\nn}{\nonumber}

\newcommand{\be}{\begin{equation}}
\newcommand{\eeq}{\end{equation}}
\newcommand{\ee}{\end{equation}}
\newcommand{\beqa}{\begin{eqnarray}}
\newcommand{\eeqa}{\end{eqnarray}}
\newcommand{\beqan}{\begin{eqnarray*}}
\newcommand{\eeqan}{\end{eqnarray*}}
\newcommand{\ba}{\begin{array}}
\newcommand{\ea}{\end{array}}

\newcommand{\mnote}[1]%{}
{\protect{\stepcounter{mnotecount}}$^{\mbox{\footnotesize
$%\!\!\!\!\!\!\,
\bullet$\themnotecount}}$ \marginpar{%\color{red}%
\raggedright\tiny\em
$\!\!\!\!\!\!\,\bullet$\themnotecount: #1} }

\newcommand{\warn}[1]%{}%{}
{\protect{\stepcounter{mnotecount}}$^{\mbox{\footnotesize
$%\!\!\!\!\!\!\,
\bullet$\themnotecount}}$ \marginpar{%\color{red}%
\raggedright\tiny\em
$\!\!\!\!\!\!\,\bullet$\themnotecount: {\bf Warning:} #1} }

\newcommand{\R}{\mathbb R}

\newcommand{\N}{\mathbb N}
\newcommand{\Z}{\mathbb Z}

\newcommand{\eq}[1]{(\ref{#1})}

\newcommand{\ptc}[1]{\mnote{{\color{blue} #1}}}

\newcommand{\beqar}{\begin{deqarr}}
\newcommand{\eeqar}{\end{deqarr}}

\newcommand{\beaa}{\begin{eqnarray*}}
\newcommand{\eeaa}{\end{eqnarray*}}

\newcommand{\bethm}{\begin{theorem}}
\newcommand{\et}{\end{theorem}}
\newcommand{\bl}{\begin{Lemma}}
\newcommand{\dnotdelta}{\mathrm{d}}

\newcommand{\notag}{\nonumber}
\def\ben{\begin{equation}}
\def\een{\end{equation}}
\def\bena{\begin{eqnarray}}
\def\eena{\end{eqnarray}}

\def\f(#1/#2){\frac{#1}{#2}}
\def\Frac(#1/#2){\left(\frac{#1}{#2}\right)}
\def\chris(#1-#2-#3){{\mit \Gamma}^{#1}{}_{{#2}{#3}} }
\def\tilchris(#1-#2-#3){\tilde{{\mit \Gamma}}^{#1}{}_{{#2}{#3}}}
\def\hatchris(#1-#2-#3){\hat{{\mit \Gamma}}^{#1}{}_{{#2}{#3}}}

\newtheorem{Theorem} {\sc  Theorem\rm} [section]
\newtheorem{theorem} [Theorem] {\sc  Theorem\rm}

\newtheorem{Lemma} [Theorem] {\sc  Lemma\rm}

\theorembodyfont{\upshape}

\theoremsymbol{\ensuremath{\diamondsuit}}
\newcommand{\fcoco}{\small}
\theorembodyfont{\fcoco}\theoremseparator{}\theoremindent0.5cm

\theoremstyle{nonumberplain}\theorembodyfont{\fcoco}
\theoremseparator{}\theoremindent0.5cm

\DeclareFontFamily{OT1}{rsfs}{}
\DeclareFontShape{OT1}{rsfs}{m}{n}{ <-7> rsfs5 <7-10> rsfs7 <10-> rsfs10}{}

\begin{document}
%\frontmatter
\title{The Euclidean quantisation of Kerr-Newman-de Sitter black holes\protect\thanks{Preprint UWThPh-2015-32}}

\author{Piotr T.\ Chru\'{s}ciel%
\thanks{ Email \protect\url{piotr.chrusciel@univie.ac.at}, URL \protect\url{http://homepage.univie.ac.at/piotr.chrusciel}}
\ and Michael H\"orzinger\\ Erwin Schr\"odinger Institute and
Faculty of Physics \\
University of Vienna }

\maketitle

\abstract{We study the family of Einstein-Maxwell instantons associated with the Kerr-Newman metrics with a positive cosmological constant. This leads to a quantisation condition on the masses, charges, and angular momentum parameters of the resulting Euclidean solutions.}
\tableofcontents

\section{Introduction}
 \label{s8IX15.1}

Euclidean counterparts of Lorentzian  solutions  play an important role in Euclidean Quantum Gravity~\cite{HawkingInstantons,EQG}. It appears therefore of interest to find Euclidean versions of key Lorentzian solutions.

As such, Kerr-Newman solutions have a unique position in view of their uniqueness properties. The associated solutions with positive cosmological constant, discovered by Demia\'nski and Pleba\'nski~\cite{PlebanskiDemianski} and, independently, by Carter~\cite{Carterseparable}, are similarly expected to be unique under natural conditions. Surprisingly enough, their compact Euclidean counterparts do not seem to have been explored in the literature. The object of this paper is to fill this gap.

More precisely, we construct two new families of compact Riemannian four-dimensional manifolds satisfying the Einstein-Maxwell equations with a positive cosmological constant. The solutions are obtained by complex substitutions in the Kerr-Newman de Sitter metric. The requirement of smoothness and compactness of the underlying manifold leads to a quantisation condition on the mass and charge parameters of the associated Lorentzian manifold. We thus obtain our first family of metrics, on $S^2$- and $\RPtwo $-bundles
 over $S^2$,
parameterised by two integers $(n_1,n_2)$. The second family is parameterised by a single integer $n\in \N$ and is obtained by passing to a limit \emph{\`a la Page} in the Euclidean Kerr-Newman de Sitter metrics.
 We determine several physical parameters associated with the Lorentzian equivalents of the solutions and study their asymptotics as one, or both, parameters tend to infinity. We calculate the associated Euclidean actions, which determine the contribution of our instantons to the Euclidean path integral in a saddle point approximation, as well as horizon entropies and temperatures.

Our Riemannian solutions $(^4M,g)$ have a clear quantum relevance. On a more mundane level, since the Maxwell energy-momentum tensor has vanishing trace,
 the metrics we have constructed provide time-symmetric initial data for the $4+1$ vacuum Einstein equations with a positive cosmological constant, or for Einstein equations with matter (e.g., dust) having constant density on the initial data surface $^4M$.
Indeed, the four-dimensional Euclidean Einstein-Maxwell equations imply that the four-dimensional Riemannian metric $g$  has constant positive scalar curvature. Therefore the initial data set $(^4 M,g, K=0)$ satisfies the $4+1$ vacuum time-symmetric constraint equations with a positive cosmological constant, or $4+1$  time-symmetric constraint equations with dust which has constant density, or with a constant scalar field, or with a mixture of the above.

The solutions in our first family are uniquely parameterized by the already mentioned quantum numbers  $(n_1,n_2)\in \N^2$, $1\le n_2 < n_1$, and the value of the cosmological constant $\Lambda$.
It might be viewed as amusing, and perhaps not entirely unexpected, that after inserting the experimentally determined value of $\Lambda$, the masses of all  Lorentzian solutions associated with our Euclidean ones are of the same order as some standard current estimates, based on the FLRW model, for the total mass of the visible universe.

The quantum numbers $(n_1,n_2)$, resulting from the requirement of regularity of the Riemannian manifold, lead  to a quantisation of the mass, the angular momentum, and the combination $p^2-e^2$ of the magnetic charge parameter $p$ and electric charge parameter $e$. We show %in Section~\ref{s1XII15.1} 
that the   requirement of a well-defined test Dirac field with charge $q_0$ on the Riemannian manifold introduces two further quantum numbers $(\hat n_1, \hat n_2)$, together with a quantisation of $e$, $p$ and $q_0$.

\section{The fields}
 \label{s1I16.1}

The  Kerr-Newman-de Sitter (KNdS) metric is a solution of the  Einstein-Maxwell equations,
\bena
R_{ \mu \nu}-\frac{1}{2} g_{ \mu \nu} R +\Lambda g_{\mu\nu}
= 8 \pi T_{ \mu \nu}\,,~~~dF=0 \,,~~~d \star F=0\,, \label{15III15.1}
\eena
where $\Lambda$ is the cosmological constant (which we assume to be positive throughout this work), and where
\bena
T_{\mu \nu}=\frac{1}{4 \pi} ( F_{\mu \rho } F _{\nu}^{\phantom{\nu}\rho } - \frac{1}{4}F^{\alpha \beta} F_{\alpha \beta} g_{\mu \nu} )\,.\label{15III15.2}
\eena
In Boyer-Lindquist coordinates, after the replacement $a \rightarrow i a $, $ t \rightarrow it$ and
$e \rightarrow i e $ the metric takes the form%
\footnote{In geometric considerations below it is convenient to scale the objects involved so that all coordinates, as well as $a$, $M$, $e$, $p$ and $\Lambda$ are unitless. When translating back to SI units in the Lorentzian metric, it is   useful to observe that $\Delta_r/r^2$ has no dimensions. Thus, if $r$ is instead measured in meters  then $\lambda r^2$, which is one of the summands of $\Delta_r/r^2$, must have no dimension and thus $\lambda$ must have dimension $m^{-2}$, etc.}

\bena
g&=& \frac{\Sigma}{\Delta_r} dr^2 +  \frac{\Sigma}{\Delta_\theta} d\theta^2 + \frac{\sin^2 (\theta)}{\Xi^2 \Sigma} \Delta_\theta ( a\dnotdelta t + (r^2 -a^2) \dnotdelta \varphi)^2 \nonumber \\
 &\phantom{=}&+ \frac{1}{\Xi^2 \Sigma} \Delta_r (dt - a \sin^2(\theta) \dnotdelta \varphi)^2\,,
\label{15III15.9}
\eena
where, setting $\lambda = \Lambda/3$,
\bena
 &
\Sigma=r^2-a^2 \cos^2(\theta) \,, ~~ \Delta_r=(r^2-a^2)\left( 1- \lambda r^2 \right) -2 M  r +p^2 - e^2 \,,
&
\label{15III18.2}
\\
 &
\Delta_\theta = 1 - \lambda a^2 \cos^2(\theta)\,,~~ \Xi=1 - \lambda a^2\,.
 &
 \label{15III18.3}
\eena
The Maxwell potential reads
\bena
	A =  \frac{p \, \cos(\theta)}{\Sigma} \sigma_1 + \frac{e\,r}{\Sigma} \sigma_2
\,,
\label{15III15.7}
\eena
where the one-forms $\sigma_i$, $i=1,2$, are defined as
\bena
	\sigma_1= \frac{1}{\Xi} \left(- a \,\dnotdelta t - (r^2-a^2) \dnotdelta \varphi \right)
\,,
\quad
	\sigma_2= \frac{1}{\Xi} \left(-\dnotdelta t + a  \sin^2(\theta) \dnotdelta \varphi \right)
\,.
\eena

Now, each  metric \eq{15III15.9} is determined uniquely by the parameters $a$, $M$, and the combination
\bel{8IX15.5}
 \esquare:= p^2 - e^2
 \,.
\ee
of the magnetic charge parameter $p$ and the electric charge parameter $e$.
The notation in \eq{8IX15.5} might appear to be misleading, because the right-hand side of this equation could be negative.  However, it turns out to be mostly appropriate, in that we have not found any non-singular solutions with $  p^2 \le e^2$ using our procedure below except in the Page limit discussed in Appendix~\ref{s7IX15.1}.

We emphasise that \emph{any} pairs $(e, p)$ satisfying \eq{8IX15.5} are allowed.
When transforming back to the Lorentzian regime, there is no ambiguity in determining the parameters $M$ and $a$ characterising the Lorentzian solution, which remain unchanged. On the other hand, if we denote by $p_L$ and $e_L$ the parameters characterising the Maxwell field on the Lorentzian side, then \emph{any} values of $p_L$ and $e_L$ satisfying
\bel{8IX15.5+}
   p^2_L + e^2_L = p^2 + e^2
\ee
are compatible with the Einstein-Maxwell equations for the \emph{Lorentzian metric}. The question thus arises  whether, given a set $(a,M,e,p)$ arising from a Riemannian metric, there is a preferred choice of $p_L$ and $e_L$.

A natural choice is
\bel{2I16.1a}
 p_L=p\,,
  \quad
  e_L=e
  \,.
\ee
The condition $\noQ>0$  and \eq{8IX15.5} imply that the simplest choice $p_L=0$ in \eq{2I16.1a} is not possible, except in the Page limit.
The next simplest choice, $e_L=0$,    leads then to purely magnetic solutions with a quantised magnetic charge.
We emphasise that our quantisation mechanism of magnetic charge has nothing to do with the Dirac one, see Section~\ref{s1XII15.1} below.

Whether or not \eq{2I16.1a} is the right choice appears to be a matter of debate, see~\cite{DunajskiTod1,HawkingRoss}.
An alternative  would be to decree that the Lorentzian solutions with $p_L=0$ and $e_L\ne 0$ correspond to Riemannian solutions  for which $\hat A :=   { e_L\,r} \sigma_2/{\Sigma}$  is a vector potential for
\bena
\star F_{\mu \nu}:= \frac{1}{2} \epsilon_{\mu \nu}^{\phantom{\mu \nu} \alpha \beta} F_{\alpha \beta}= \partial_\mu \hat A_\nu -\partial_\nu \hat A_\mu \,,
\eena
where $ \epsilon_{\mu \nu \alpha \beta}$ is the totally antisymmetric tensor. In this case $\esquare=e^2_L$ (compare~\cite{DunajskiTod1}). This choice leads to a quantisation of electric charge.

It might be of interest to note that \emph{planar Lorentz transformations} of $(p,e)$  preserve $\esquare$, and can be thought of as the Euclidean counterparts of the usual duality transformations of the Maxwell field, which instead act as rotations of the $(p,e)$ plane.

In any case, we wish to find ranges of parameters so that \eq{15III15.9} is a Riemannian metric on a closed manifold $M$. This leads to the following obvious restrictions:

First, compactness requires $\varphi$ and $t$ to be periodic, with a period which needs to be determined.

Further, compactness of $M$ requires a  range of the variable $r$, bounded by two \emph{first-order} zeros $r_1<r_2$ of $\Delta_r$, so that (\ref{15III15.9}) is Riemannian for $\forall r \in (r_1,r_2),\;\theta \in (0,\pi)$.%
\footnote{One can likewise enquire about existence of compact Euclidean solutions with $\Lambda\le0$. One easily checks that for $\Lambda\le 0$ the function $\Delta_r$ has no maxima in the range of parameters of interest, and therefore no configurations as considered here exist.}
In particular
\bena
\frac{\Sigma}{\Delta_r} >0 ~~~\text{and}~~~\frac{\Sigma}{\Delta_\theta}>0~~\forall r \in (r_1,r_2),\;\theta \in (0,\pi).
\eena
Equations~(\ref{15III18.2}) and (\ref{15III18.3}) show that $\Sigma$ and    $ \Delta_\theta  $ are positive on the equatorial plane, and we conclude that
\bena
\Delta_r>0\,,~~~\Sigma>0~~~\text{and}~~~\Delta_\theta>0~~~\forall r \in (r_1,r_2),\;\theta \in (0,\pi)\,.
\eena
Now, if $r_1 r_2\le 0$, then $0\in [r_1,r_2]$, and since $\Sigma|_{r=0}<0$ this case will not lead to a regular Riemannian metric. Changing $r$ to its negative, it remains to consider the case where $0<  r_1 <r_2$. Positivity of $\Sigma$ leads then to $r_1>|a|$, and positivity of $\Delta_\theta$ imposes the restriction $\lambda^{-1}>a^2$.
 Summarising:
\bena
0<|a|<r_1<r_2
\,,
 \quad a^2< \lambda^{-1}
\,,
 \quad
  \Delta_r|_{(r_1,r_2)}>0
 \,.
\label{3V15.1}
\eena

Given a Euclidean metric as above \emph{with $e=0$}, the corresponding Lorentzian metric with the same real values of $\lambda$, $M$, $a$, $e=0$, and $p$ will be called a \emph{partner solution}.
Note that  the locations $r_i$ of the horizons of the partner solution will \emph{not} coincide with the locations $r_i$ of the rotation axes of the associated Euclidean solutions; similarly for areas, surface gravities, etc.
%%%%%%%%%%%%%%%%%%%%%%%%%%%%%%%%%%%%%%%%%%%%%%%%%%%%%%%%%%%%%%%%%%%%%%%%%%%%%%%%%%%%%%%%%%%%%%%

%%%\input{thetaAxesRegularity_corrCH}%%%%%%%%%%%%%%%%%%%%%%%%%%%%%%%%%%%%%%%%%%%%%%%%%%%%%%%%%%

\section{Regularity at the rotation axes}
 \label{s2V15.1}

For $r\in[r_1,r_2]$ let us introduce two functions $\rho_i$,   $i=1,2$, defined as
\bena
\rho_i= \epsilon_i\int_{r_i}^r \frac{1}{\sqrt{\Delta_r}}  \dnotdelta r
  = \frac{2}{  \sqrt{ \almostkappa_{i} }}\sqrt{\epsilon_i(r-r_i)}\mathbb{1}_{1,i}(r-r_i)\,,
\eena
with $\epsilon_1=1$, $\epsilon_2=-1$, and
\bel{10V15.11}
 \almostkappa_{i}:=  \left|\Delta_{r} ' |_{r={r_i}}\right| \neq 0
 \,,
 \quad  i=1,2
 \,,
\ee
and with functions $\mathbb{1}_{1,i}$ which are smooth near the origin and satisfy $\mathbb{1}_{1,i}(0)=1$. The function $\rho_1$ will serve as a coordinate replacing $r$ for $r\in[r_1,r_2)$, while $\rho_2$ will replace $r$ for $r\in(r_1,r_2]$.
Inverting, it follows that
\bena
r=r_1 + \frac{\almostkappa_{1}}{4} \rho_1^2 \mathbb{1}_2(\rho_1^2)
 \,,
 \qquad
 \Delta_r =  \frac{\almostkappa_{1}^2}{4}  \rho_1^2 \mathbb{1}_3(\rho_1^2)
 \,,
\eena
with functions $\mathbb{1}_2$, $\mathbb{1}_3$ which are smooth near the origin, with $\mathbb{1}_2(0)=1=\mathbb{1}_3(0)$.

In order to make sure that the metric is regular near the intersection of the axes $\{\sin \theta=0\}$ with the axes $\{\Delta_r=0\}$, near $\theta=0$ and for
$r\in[r_1,r_2)$ we use a coordinate system
$(\rho_1,t_1,\theta,\phi_1)$,
with $t=\omega_1 t_1$  and $\varphi$ defined through the formula
\bel{5V15.1}
 \dnotdelta \varphi
  := \alpha_1 \dnotdelta\phi_1 +  \frac{a  }{ a^2-r_1^2 \color{black}} \dnotdelta t
  \equiv \alpha_1 \dnotdelta\phi_1 +  \frac{a  \omega_1}{ a^2-r_1^2 \color{black}} \dnotdelta t_1
 \,,
\ee
for some constants $\alpha_1,\omega_1 \in \R^*$ which will be determined shortly by requiring $2\pi$-periodicity of $t_1$ and $\phi_1$. In \eq{5V15.1}   the coefficient in front of $ \dnotdelta t $ has been chosen so that $g_{t t }|_{\rho_1=0}=0$.
In this coordinate system the metric takes the form
\bean
 g
  &  = &
   {\Sigma} \bigg\{
    \dnotdelta \rho_1^2
    + \frac{1}{\Xi^2 \Sigma ^2}
        \bigg[ \frac{   \almostkappa_1^2 \omega_1^2 \Sigma ^2}{4\left( r_1^2-a^2 \color{black} \right)^2} \mathbb{1}_4(\rho_1 ^2,\sin^2(\theta)) \rho_1 ^2 \dnotdelta t_1^2
\\
 \nonumber
  &&
   \phantom{ xx}
        + \alpha_1 ^2 \big(\Delta_\theta \left(a^2-r ^2\right)^2
         +a^2 \Delta_r \sin ^2(\theta ) \big)
         \sin^2(\theta) \dnotdelta \phi_1^2
\\
% \nonumber
  &&
   \phantom{ xx}
 + F(\rho^2_1,\sin^2(\theta))
   \rho_1 ^2 \sin   ^2(\theta )
   \dnotdelta t_1 \dnotdelta \phi_1
    \bigg]
% \nonumber
%\\
% &&
 +  \frac{1}{\Delta_\theta} \dnotdelta  \theta^2
 \bigg\}
 \,,
\eeal{6V15.2}
for some smooth functions $\mathbb{1}_4$ and $F$, with $\mathbb{1}_4(0,y)=1$.
As is well known, when $(\rho_1,t_1)$ are viewed as polar coordinates around $\rho_1=0$, the one form $\rho_1^2 \dnotdelta t_1$ and the quadratic form $\dnotdelta\rho_1^2 + \rho_1^2 \dnotdelta t_1^2$ are smooth.
Similarly when $(\theta,\phi_1)$ are  polar coordinates around $ \theta=0$, the one form
$\sin^2(\theta)\dnotdelta \phi_1$ and the quadratic form $\dnotdelta\theta^2 + \sin(\theta)^ 2\dnotdelta \phi_1^2$ are smooth. It is then easily inferred that
the requirements of $2\pi$-periodicity of $  t_1$ and $\phi_1$, together with
\bel{7V15.2}
 \frac{   \almostkappa_1^2 \omega^2_1 }{4\Xi^2\left( r_1^2-a^2 \color{black}\right)^2} =1
\,,
\quad
 \frac{ \alpha_1 ^2 \Delta_\theta^2 \left(a^2-r ^2\right)^2 }{\Xi^2 (r^2-a^2 \cos^2(\theta))^2}\bigg|_{  \theta =0} = 1
 \,,
\ee
implies smoothness both of  the sum of the diagonal terms of the metric $g$  and of the off-diagonal term $g_{t_1\phi_1} \dnotdelta t_1 \dnotdelta \phi_1$  on
$$
 \{(r,t_1,\theta,\phi_1)\in[r_1,r_2)\times S^1\times [0,\pi)\times S^1\}
 \,.
$$
%.
Note that \eq{7V15.2} is equivalent to
\bel{7V15.4}
\omega_1 = \pm \underbrace{\frac{ 2 \Xi \left(r_1^2-a^2\right) }{  \almostkappa_1}}_{=:\omega}
\,,
\quad
 \alpha_1   = \pm 1
 \,.
\ee

The above calculations remain valid without changes near $\theta=\pi$. It is, however, convenient, to use a different symbol for the resulting polar coordinates: When $\theta\in(0,\pi]$ we will use $\hat t_1$ and $\hat \phi_1$ for the relevant angular coordinates, and $\hat \omega_1$, $\hat \alpha_1$ for the corresponding coefficients. Thus,  for $\theta\in(0,\pi]$:
\bel{15V15.51}
 t=\hat \omega_1 \hat t_1\,,
 \quad
  \dnotdelta \varphi= \hat \alpha_1 \dnotdelta\hat \phi_1 +  \frac{a \hat \omega_1}{ a^2-r_1^2 \color{black}} \dnotdelta \hat t_1
 \,,
\ee
with
\bel{15V15.52}
\hat \omega_1 = \pm \omega
\,,
\quad
 \hat \alpha_1   = \pm 1
 \,.
\ee

Identical considerations  for $r\in (r_1,r_2]$, using  coordinate systems
$(\rho_2,t_2 = t \omega_2^{-1},\theta,\phi_2)$ for $\theta\in[0,\pi)$ and
$(\rho_2,\hat t_2 = t \hat \omega_2^{-1},\theta,\hat \phi_2)$ for $\theta\in(0,\pi]$,
with
\bel{7V15.3}
 \dnotdelta \varphi
  = \alpha_2 \dnotdelta\phi_2 +  \frac{a \omega_2}{a^2-r_2^2 \color{black}}  \dnotdelta t_2
 \,,
 \qquad
 \dnotdelta \varphi
  = \hat \alpha_2 \dnotdelta\hat \phi_2 +  \frac{a \hat  \omega_2}{ a^2-r_2^2 \color{black}}  \dnotdelta \hat t_2
 \,,
\ee
lead to
\bel{7V15.5}
  \omega_2\,, \,    \hat \omega_2 \in \left\{ \pm \frac{2 \Xi \left(r_2^2-a^2\right) }{ \almostkappa_2}\right\}
\,,
\quad
 \alpha_2\,,\, \hat \alpha_2   \in\{ \pm 1 \}
 \,.
\ee

In an overlap region where both $t$ and $t_1$ are coordinates, the equation $t=\omega_1t_1$ implies that $t$ must be exactly $2\pi|\omega_1|$-periodic. Similarly, in any overlap region where both $t$ and $t_2$ are defined and are coordinates,  $t$ must be exactly $2\pi|\omega_2|$-periodic. A similar argument applies to $\hat t_i$. So, the periodicity requirements of $t_i$ and $\hat t_i$ lead to
\bel{8V15.1}
 \omega_1 = \pm \omega_2
 \,,
 \qquad
 \hat \omega_1 = \pm \hat \omega_2
 \,.
\ee
\subsection{$a=0$}
 \label{ss20XI15.1}
 When $a=0$, and imposing the regularity conditions above, the metric \eq{6V15.2asdf} simplifies considerably:
\bea
 g
  &  = &
   {r^2} \bigg\{
    \dnotdelta \rho_1^2
    +    \mathbb{1}_4(\rho_1 ^2,\sin^2(\theta)) \rho_1 ^2 \dnotdelta t_1^2
 +    \dnotdelta  \theta^2
        +    \sin^2(\theta) \dnotdelta \phi_1^2
 \bigg\}
 \,.
\eeal{6V15.2asdf}
The coordinate $\rho_1$ can be written explicitly in terms of elliptic integrals, which is not very enlightening.

\begin{figure}
\begin{center}
%\resizebox{1.5in}{!}
{\includegraphics[scale=.6]{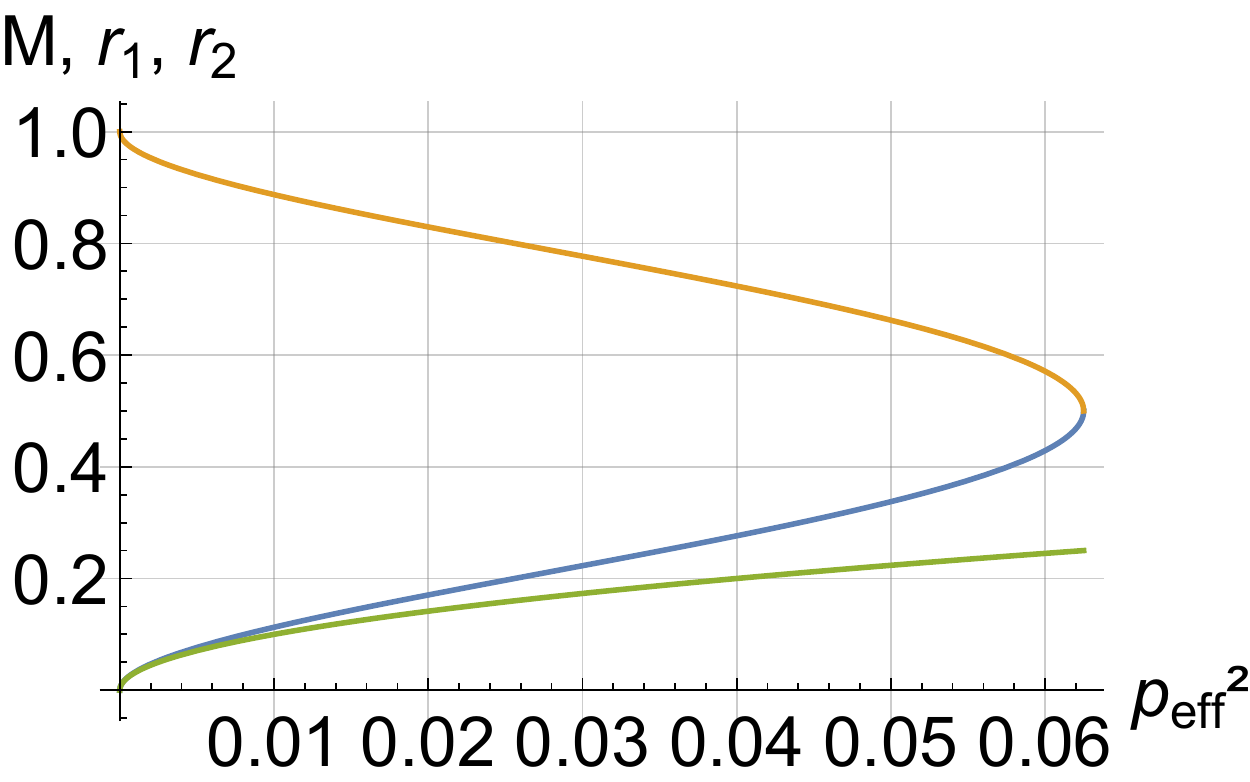}}
\caption{Solutions with $a=0$ scaled to $\lambda=1$. The uppermost curve is a plot of $r_2$, the middle one that of $r_1$, the lowest curve is a plot of the mass parameter $M$.}
\label{F20XI15.1}
\end{center}
\end{figure}

After scaling to $\lambda=1$,
the periodicity conditions \eq{8V15.1} are verified by a one-parameter family of solutions parameterized by a continuous parameter $\noQ\in[0,1/16)$,
see Figure~\ref{F20XI15.1}. These solutions will not be discussed any further.

\subsection{$ {a\ne 0}$: the quantisation conditions}

When $a\ne 0$, without loss of generality, replacing $t$ and/or $\varphi$ by their negatives if necessary,  we require
\bel{21VI15.21}
 a>0\,,
 \quad
  \omega_1=\omega >0
  \,.
\ee

To avoid ambiguities: except for the analysis of the Page limit in Appendix~\ref{s7IX15.1}, in what follows we will assume that \emph{$(r,t,\theta,\varphi)$ form a smooth coordinate system  away from the rotation axes}, with $t$ and $\varphi$ periodic.

Increasing $\phi_1$ from zero to $2\pi$ with $(\rho_1,t_1,\theta)$ fixed takes one back to the starting point.
Equations~\eq{5V15.1} and \eq{7V15.4} show that $\varphi$ changes by $\pm 2 \pi$, and therefore the minimal period of $\varphi$ must be  $2\pi/k$ for some $k\in \N^*$. But then, increasing $\varphi$ from zero to $2\pi/k$ with $(r,t,\theta)$ fixed takes one to the same place. This results in an increase of $\phi_1$ by $\pm 2\pi/k$, which implies that $k=1$. Hence, $\varphi$ is exactly $2\pi$-periodic.

Now, increasing $t_1$ from zero to $2 \pi$ with $\varphi_1$ fixed again takes one to the same place. This implies that $\varphi$ must have changed by an integer multiple of $2\pi$.

The same argument applies to $\varphi_2$ and $t_2$. We conclude that
%there exist integers $n_i $ such that
%
\bel{15V15.21}
n_i:= \frac{a \omega }{r_i^2-a^2} \in \N^{*}
 \,.
\ee

\subsection{Maxwell fields}
 \label{ss1XII15.1}

Let us check that the Maxwell fields, defined as $\dnotdelta A$ \emph{away from all axes of rotation}, extend by continuity to smooth fields once the above constraints have been imposed. This can be done by inspection of the Maxwell potentials \eq{15III15.7} (which, incidentally, are \emph{not} regular at the rotation axes).

We start with an analysis of the $p$-contribution to $A$ which, using \eq{5V15.1} and its equivalent with $r_1$ replaced by $r_2$, can be rewritten as
\bena
    \nn
    \lefteqn{
	 \frac{p \, \cos(\theta)}{\Sigma} \sigma_1
   =\frac{p \, \cos(\theta)}{\Xi\Sigma } \left(- a \,\dnotdelta t - (r^2-a^2) \dnotdelta \varphi \right)
    }
    &&
\\
\nn
 &   & =
   \frac{p \, \cos(\theta)}{\Xi\Sigma } \left(- a \,\dnotdelta t - (r^2-a^2)
 \left(
  \alpha_i \dnotdelta\phi_i +  \frac{a  }{ a^2-r_i^2 \color{black}} \dnotdelta t
  \right) \right)
\\
\nn
 &   & =
    \frac{p \, \cos(\theta)}{\Xi\Sigma } \left(\frac{ a (r^2-r_i^2)}{r_i^2-a^2 } \,\dnotdelta t - \alpha_i(r^2-a^2)
 \dnotdelta\phi_i  \right)
\\
 &   & =
    \underbrace{\frac{p \, \cos(\theta)}{\Xi} \left(\frac{ a (r^2-r_i^2)}{\Sigma (r_i^2-a^2) } \,\dnotdelta t + \frac{\alpha_i a^2\sin^2\theta}{r^2-a^2\cos^2(\theta)}
 \dnotdelta\phi_i  \right)}_{\text{\rm smooth}}
  -  \frac{\alpha_i p \, \cos(\theta)}{\Xi} \dnotdelta\phi_i
\,,
\label{15III15.7x}
\eena
where the index $i$ on $r_i$, $\alpha_i$ and $\phi_i$ takes the values $i\in\{1,2\}$. More precisely, the underbraced term in the last line of \eq{15III15.7x} is smooth away from $r=r_2$ when $i=1$, and away from $r=r_1$ when $i=2$.
Near the axis $\cos \theta = 1$ the last, \emph{non-smooth} term can be rewritten as
\bel{1XII15.11}
  -  \frac{\alpha_i p \, \cos(\theta)}{\Xi} \dnotdelta\phi_i =
  -  \underbrace{\frac{\alpha_i p \, (\cos(\theta)-1)}{\Xi} \dnotdelta\phi_i}_{\text{\rm smooth for $\theta<\pi$}} -
    \underbrace{\frac{\alpha_i p  }{\Xi} \dnotdelta\phi_i }_{\mathrm{closed}}
    \,,
\ee
which shows smoothness of the $p$-contribution to $F=\dnotdelta A$ near $
\theta=0$.

In fact, we have proved that, for $i=1,2$, the vector potentials
\bel{1I16.4}
 \frac{p \, \cos(\theta)}{\Sigma} \sigma_1 + \frac{\alpha_i p  }{\Xi} \dnotdelta\phi_i
 =
  \frac{p \, \cos(\theta)}{\Sigma} \sigma_1 + \frac{ p  }{\Xi}  ( \dnotdelta \varphi
 - \frac{a  }{ a^2-r_i^2 } \dnotdelta t)
\ee
which are well-defined and smooth away from all axes of rotation, extend  by continuity across $\theta=0$ and $r=r_i$ to smooth covector fields.

An identical calculation
near the axis $\cos \theta = - 1$, with $\phi_i$ replaced by $\hat \phi_i$, shows that  the offending  term can be rewritten as
\bel{1XII15.12}
  -  \frac{\alpha_i p \, \cos(\theta)}{\Xi} \dnotdelta\hat \phi_i =
  -  \underbrace{\frac{\alpha_i p \, (\cos(\theta)+1)}{\Xi} \dnotdelta\hat \phi_i}_{\text{\rm smooth for $\theta>0$}} +
    \underbrace{\frac{\alpha_i p  }{\Xi} \dnotdelta\hat \phi_i }_{\mathrm{closed}}
    \,,
\ee
which finishes the proof of smoothness of the $p$-contribution to $F $ everywhere.
We also see that the potentials
\bel{1I16.4+}
  \frac{p \, \cos(\theta)}{\Sigma} \sigma_1 - \frac{\alpha_i p  }{\Xi} \dnotdelta\hat \phi_i
 =
  \frac{p \, \cos(\theta)}{\Sigma} \sigma_1 - \frac{ p  }{\Xi}  ( \dnotdelta \varphi
 - \frac{a  }{ a^2-r_i^2 } \dnotdelta t)
\ee
extend smoothly to the axis $\theta=\pi$.

We continue with the $e$-contribution to $A$:
\bena
 \nn
 \lefteqn{
   \frac{e\,r}{\Sigma} \sigma_2
   =
    \frac{e\,r}{\Sigma\Xi}
    \left(-\dnotdelta t + a  \sin^2(\theta) \dnotdelta \varphi \right)
   =
    \frac{e\,r}{\Sigma\Xi}
    \left(-\dnotdelta t + a  \sin^2(\theta)\left(
    \alpha_i \dnotdelta\phi_i +  \frac{a  }{ a^2-r_i^2 } \dnotdelta t
    \right)
     \right)
    }
    &&
\\
 \nn
 & =&
    \frac{e\,r}{\Sigma\Xi}
    \left(  \frac{ r_i^2 -a^2 \cos^2(\theta) }{ a^2-r_i^2 } \dnotdelta t + a \alpha_i \sin^2(\theta)
    \dnotdelta\phi_i
     \right)
\\
 \nn
 & =&
     \frac{e }{ \Xi}
    \left(  \frac{ r_i^2 -a^2 \cos^2(\theta) }{   r ^2 -a^2 \cos^2(\theta) } \times \frac{ r}{(a^2-r_i^2) } \dnotdelta t + \frac{\alpha_i a r \sin^2(\theta)}{  r ^2 -a^2 \cos^2(\theta) }
    \dnotdelta\phi_i
     \right)
\\
 \nn
 & =&
     \frac{e }{ \Xi}
    \left( \left(1- \frac{r^2- r_i^2   }{   r ^2 -a^2 \cos^2(\theta) } \right) \frac{ r}{(a^2-r_i^2) } \dnotdelta t + \frac{\alpha_i a r \sin^2(\theta)}{  r ^2 -a^2 \cos^2(\theta) }
    \dnotdelta\phi_i
     \right)
\\
 \nn
 & =&
   \underbrace{
    \frac{e }{ \Xi}
    \left(  \frac{ r- r_i }{(a^2-r_i^2) } \dnotdelta t
    -  \frac{r^2- r_i^2   }{   r ^2 -a^2 \cos^2(\theta) } \times  \frac{ r}{(a^2-r_i^2) } \dnotdelta t +
     \frac{\alpha_i a r \sin^2(\theta)}{  r ^2 -a^2 \cos^2(\theta) }
    \dnotdelta\phi_i
     \right)
     }_{\text{\rm smooth near $r=r_i$}}
\\
 && + \underbrace{\frac{e r_i }{ \Xi (a^2-r_i^2) } \dnotdelta t}_{\text{\rm closed}}
 \,.
\label{1XII15.1}
\eena
This finishes the proof of smoothness of $F$.

Note that \eq{1XII15.1} shows that $ {e\,r}\,\sigma_2/\Sigma $ extends smoothly across both $\theta=0$ and $\theta=\pi$ without further due as long as one stays away from the axes $r\in\{r_1,r_2\}$.

\section{Topology}
 \label{s21VI15.2}

The results of Section~\ref{s2V15.1}
can be summarised as follows: imposing $2\pi$-periodicity of $t_1$, $\hat t_1$, $t_2$, $\hat t_2$ $\varphi_1$, $\hat \varphi_1$, $\varphi_2$ and $\hat \varphi_2$, together with $\omega_1= \omega$, and $\hat \omega_1\,,\, \omega_2\,,\, \hat  \omega_2\in \{\pm \omega\}$, as well as $\alpha_1$, $\hat \alpha_1\,,\, \alpha_2\,,\, \hat \alpha_2\in \{\pm 1\}$ and \eq{15V15.21}, the coordinates $(\rho_i,t_i,\theta,\varphi_i)$,  $(\rho_i,\hat t_i,\theta,\hat \varphi_i)$,   $i=1,2$ such that
\beal{13V15.1-}
&
\rho_i (r)
   =
  \int_{r_i}^r \frac{1}{\sqrt{\Delta_r}}  \dnotdelta r
  \,,
  \qquad
 \omega t_1 = t =\pm \omega t_2
  \,,
  &
\\
 &
    \alpha_1 \dnotdelta\phi_1 +   \frac{a \omega }{r_1^2-a^2} \dnotdelta t_1
    =
 \dnotdelta \varphi
    =
   \alpha_2 \dnotdelta\phi_2   \pm \frac{a \omega }{r_2^2-a^2} \dnotdelta t_2
 \,,
 &
\eeal{13V15.1}
similarly for the hatted ones,
provide polar coordinates on the following four distinct coordinate patches, each containing exactly one intersection of the axes of rotation $\{\Delta_r=0\}\cap \{\sin (\theta)=0\}$ in their centers:
\beal{21VI15.11}
 \Omega_{r_1,0}
  & := &
   [r_1,r_2)_{\rho_1}\times S^1_{t_1}\times [0,  \pi )_{\theta}\times S^1_{\phi_1}
 \approx D^2_{(\rho_1,t_1)}\times D^2_{(\theta,\phi_1)}
 \,,
\\
   \Omega_{r_2,0}
    & := &
    (r_1,r_2]_{\rho_2}\times S^1_{t_2}\times [0,  \pi )_{\theta}\times S^1_{\phi_2}
 \approx D^2_{(\rho_2,t_2)}\times D^2_{(\theta,\phi_2)}
 \,,
\\
 \Omega_{r_1,\pi}
  & := &
   [r_1,r_2)_{\rho_1}\times S^1_{\hat t_1}\times (0,  \pi ]_{\theta}\times S^1_{\hat \phi_1}
 \approx D^2_{(\rho_1,\hat t_1)}\times D^2_{(\theta,\hat \phi_1)}
 \,,
\\
   \Omega_{r_2,\pi}
    & := &
    (r_1,r_2]_{\rho_2}\times S^1_{\hat t_2}\times (0,  \pi ]_{\theta}\times S^1_{\hat \phi_2}
 \approx D^2_{(\rho_2,\hat t_2)}\times D^2_{(\theta,\hat \phi_2)}
 \,.
\eeal{21VI15.14}
Here ``$\approx$'' means ``diffeomorphic to'', and  $D^2_{(\rho_1,t_1)}$ denotes an open disc  $D^2\subset \R^2$ coordinatised by polar coordinates ${(\rho_1,t_1)}$ while $ S^1_{t_2}$ denotes a circle $S^1$ coordinatised by $t_2$, etc. Quite generally, we use the notation $U_x$ to denote the fact that a set $U$ is coordinatised by a variable $x$.

The question then arises, in how many ways can one  glue the sets above to obtain  smooth closed manifolds. We point out some possible constructions here. While we suspect that these are all possibilities, we have not made in-depth attempts to analyse whether or not the list below is exhaustive.%
\footnote{The solutions we construct are $U(1)\times U(1)$-symmetric, and the results in~\cite{OrlikRaymondII} are relevant in this context. However, one could also search for manifolds carrying the metric \eq{15III15.9} which are only locally $U(1)\times U(1)$-symmetric.}
Note that oriented manifolds are obtained if and only if $\omega_2 = \alpha_1 \alpha_2 \omega$.

\begin{enumerate}
  \item
We can glue $\Omega_{r_1,0}$ with $\Omega_{r_1,\pi}$ by identifying  for $\theta\in(0,\pi)$ the points  $(\rho_1,t_1,\theta,\varphi_1)$  with $(\rho_1,\hat t_1,\theta,\hat \varphi_1)$; similarly for $\Omega_{r_2,0}$ and $\Omega_{r_2,\pi}$. This corresponds to the choice $\alpha_1=\alpha_2$, and leads to the manifolds
$$
 \widehat \Omega_1^+ :=[r_1,r_2)_{\rho_1}\times S^1_{t_1}\times [0,  \pi] _{\theta}\times S^1_{\phi_1}
 \approx D^2_{(\rho_1,t_1)}\times S^2_{(\theta,\phi_1)}
 \,,
$$
as well as
$$
    \widehat \Omega_2^+ :=(r_1,r_2]_{\rho_2}\times S^1_{t_2}\times [0,  \pi ]_{\theta}\times S^1_{\phi_2}
 \approx D^2_{(\rho_2,t_2)}\times S^2_{(\theta,\phi_2)}
 \,,
$$
where $S^2$ denotes a two-dimensional sphere.

Since the map $ \theta\to \pi-\theta$ is an isometry, a second possibility in the same spirit is to identify   for
$\theta\in(0,\pi)$ the points  $(\rho_1,t_1,\theta,\varphi_1)$  with $(\rho_1,\hat t_1,\pi-\theta,\hat \varphi_1+\pi)$. This leads to $\PRtwo$ bundles
over $ D^2_{(\rho_1,t_1)}$ and $D^2_{(\rho_2,t_2)}$, which are not orientable.

 \item  Let us set
 \bel{8IX15.1}
  \zeta := \omega_2/\omega_1\in \{\pm 1\}
  \quad \Rightarrow
  \quad
  \dnotdelta t_1 = \zeta  \dnotdelta t_2
   \,.
 \ee
 %.
 Consider the manifolds $  \widehat \Omega_i^+$, $i=1,2$. Both are trivial $S^2$ bundles over the  open disc $D^2$. Near the boundary of $D^2$, for each $t_2$ the corresponding sphere at $t_1$ is obtained by rotating $S^2$ around the $z$-axis by an angle $ \alpha_1 \zeta  (n_2- n_1)t_2$:
 \bel{25V15.1}
    \alpha_1 \dnotdelta\phi_1 +   n_1 \dnotdelta t_1
    =
    \alpha_ 2 \dnotdelta\phi_2  + \zeta  n_2 \dnotdelta t_2
   \quad
   \Longrightarrow
   \quad
   \phi_1 =   \alpha_1\alpha_2 \phi_2  +  \alpha_1 \zeta  (n_2 - n_1)   t_2 +c
 \,,
\ee
for some constant $c$.
So, as we circle around the boundary of $D^2$, the sphere $S^2$ is rotated by a total angle $2 \alpha_1\zeta  (n_2-n_1)\pi$ during each revolution. The end manifold is a non-trivial sphere bundle over $S^2$ when $n_2-n_1$ is odd.

A similar construction applies to the $\PRtwo$
bundles above.

\item
Let $(\rho_1){}_{\max}$ be the maximal value of the coordinate $\rho_1$, thus
\bean
\rho_1 (r)
   & = &
  \int_{r_i}^r \frac{1}{\sqrt{\Delta_r}}  \dnotdelta r
   =
  \underbrace{\int_{r_1i}^{r_2} \frac{1}{\sqrt{\Delta_r}}  \dnotdelta r}_{=(\rho_1){}_{\max}}-
  \underbrace{\int_r^{r_2} \frac{1}{\sqrt{\Delta_r}}  \dnotdelta r}_{=\rho_2}
\\
  & = &
    (\rho_1){}_{\max} - \rho_2
  \,,
  \qquad
\eeal{21XI15.1}
and suppose that the map
$$
 (\rho_1=\rho,t_1=s)\mapsto ( \rho_1=(\rho_1){}_{\max}-\rho , t_1=s+\pi)
$$
is an isometry.
    This, however, occurs  for the metrics considered here only in the Page limit, and is therefore only relevant to Appendix~\ref{s7IX15.1}.
 Then the identification of $(\rho_1,t_1,\theta,\phi_1)$ with
 $$
  (\rho_2=(\rho_1){}_{\max}-\rho_1,\hat t_2= t_1 + \pi, \pi-\theta , n( \phi_1+\pi))
 $$
 leads to a smooth compact manifold.
\end{enumerate}

\section{The solutions}
 \label{s21VI15.3}

The question then arises to find values of $(M,\noQ ,a)$ so that
\bena
\omega_1 = \zeta  \omega_2
 \,,
 \quad
  \zeta \in \{\pm 1\}
  \,,
 \quad
\frac{ a \omega_1}{r_1^2-a^2}=n_1 \in \N^*\,,
\quad
  \frac{ a |\omega_2|}{r_2^2-a^2}=n_2 \in \N^*\,.
\eena
It follows from (\ref{7V15.4}) and (\ref{7V15.5})  that the above equations are equivalent to
\bena
\frac{ (r_1^2-a^2)}{(r_2^2-a^2)} \frac{\Delta'_r(r_2)}{\Delta'_r(r_1)}=  -1   \,,
\quad
\Delta'_r(r_1) n_1=  2 a \Xi\,,  \quad
-  \Delta'_r(r_2) n_2 =  2 a \Xi\,.   \label{19V15.2}
\eena
In addition we need to fulfill $
\Delta_r(r_i)=0$, leading to the system of polynomial equations for $(r_1, r_2, n_1, n_2, a, M, \noQ)$.
\bena
\Delta_r(r_1)&=&0 \,, \label{15I12.0}\\
\Delta_r(r_2)&=&0 \,, \label{15I12.1}\\
 \Delta'_r(r_1)  n_1 - 2 a \Xi &=& 0\,, \label{15I12.2}\\
 -  \Delta'_r(r_2)  n_2 - 2 a \Xi &=& 0\,, \label{15I12.3}\\
(r_1^2-a^2) n_1 -   (r_2^2-a^2) n_2 &=&0 \,. \label{15I12.4}
\eena
Note that $n_1>n_2\ge 1$ in view of \eq{15I12.4}.
Moreover the solutions have to satisfy the constraints

 \bigskip

$i)\phantom{ii}~M\in \R$, $a>0$, $\noQ  \in \R$;

$ii)\phantom{i}~n_1,n_2 \in \mathbb{N}^*$;

$iii)~0<r_1<r_2$, $|a|<|r_1|$ and $a^2<\lambda^{-1}$.

\bigskip

We note that we also need  $\forall r \in (r_1,r_2) : \Delta_r(r)>0$, but this follows from the fact that $\Delta'_r(r_1)$ is positive by \eq{15I12.2} and  $\Delta'_r(r_2)$ is negative by \eq{15I12.3}.

We also note that equations \eq{15I12.0}-\eq{15I12.4}   involve neither $\zeta $ nor the $\alpha_i$'s as in \eq{8IX15.1}-\eq{25V15.1}, which can thus be arbitrarily chosen once a solution has been found.

Our strategy is to prescribe $\lambda \in \R_+^*$,$~n_1,n_2 \in \N^*$ so that (\ref{15I12.0})-(\ref{15I12.4}) become  a system of five polynomials in the variables $(r_1,r_2,\noQ ,M,a)$. We use {\sc Mathematica} to compute a Gr\"obner
basis of the system. This provides a simpler equivalent system to solve. It turns out that one is led to a hierarchic system of polynomial equations, the first one depending only on $\noQ $, the second one only on $\noQ$ and $a$, and so forth. An example is provided in Appendix~\ref{A21VI15.1}.

Our {\sc Mathematica} calculations show the following:
Let
\bel{21VI15.31}
 \nmax = 50
\,.
\ee
Then:
\begin{enumerate} \item
    There exist no solutions with   $(n_1,n_2)\in \N \times\N $ with $1\le n_2 < n_1 \le \nmax$ and $\noQ\le0$. In particular there are no vacuum solutions with the properties set forth above.
   \item
    For every pair $(n_1,n_2)\in \N \times\N $ with $1\le n_2 < n_1 \le \nmax$ there exists exactly one solution satisfying our constraints.
      \item The physical parameters (see Appendix~\ref{s31XII15.4})
       of the Lorentzian partner solutions are all bounded, cf.~Table~\ref{T8IX15.1}. In particular the physical mass of the Lorentzian
           partners is strictly positive, bounded away from zero, and bounded from above.
\end{enumerate}
 \renewcommand{\arraystretch}{1.2}
\begin{table}[h!]
\hspace{-4cm}
\begin{minipage}{0.2\textwidth}
\begin{tabular}{ | l | c | l | }
\hline  & $(n_{1 },n_{2 })_{\min{}}$ & min.
\\ \hline
$\ephysphys{}$ & $(2,1)$ & 0.2511 \\ $M_{\phys}$ & $(\infty,1)$ & 0.2036 \\  $|J_{\phys}|$ & $(2,1)$ & 0.01392  \\ $S$ & $(2,1)$ & -2.357 \\ \hline  \end{tabular}
\end{minipage}\qquad \qquad \qquad \qquad \qquad
\begin{minipage}{0.2\textwidth}
\begin{tabular}{ | l | c | l | }
\hline  & $(n_{1 },n_{2 })_{\max{}}$ & max.
\\ \hline
$\ephysphys{}$ & $(\infty,\infty)$ & $\frac{\sqrt{2}}{3} \approx 0.47$ \\ $M_{\phys}$ & $(100,90)$ & 0.2548 \\  $|J_{\phys}|$ & $(\infty,\infty)$ & $\frac{1}{9} \approx 0.111$ \\ $S$ & $(\infty, n_2)$ & $\infty$  \\ \hline  \end{tabular}
\end{minipage}
\caption{\label{T8IX15.1}
Left table: Minimal values of the effective physical Lorentzian charge $\ephysphys{}$, the physical mass $M_{\phys}$, the physical angular momentum $|J_{\phys}|$, and the Euclidean action $S$ with the corresponding quantum numbers  $(n_{1 },n_{2 })_{\min{}}$. Right table: Maximal values of $\ephysphys{}:=\sqrt{\noQ}/(1+a^2)$, $M_{\phys}$, $|J_{\phys}|$, $S$ with the corresponding quantum
numbers  $(n_{1 },n_{2 })_{\max{}}$. All values scaled to $\lambda=1$; compare Appendix~\ref{SI}.}
\end{table}

It should be emphasised that the existence of the solutions of the system as above is a rigorous result, derived by exact computer algebra. While numerics is used to check whether the joint zeros of the Gr\"obner basis satisfy the desired inequalities, this is again a rigorous statement, as the numerical errors introduced when checking the inequalities are well below the gaps occurring in the inequalities.

\begin{figure}
\begin{center}
\resizebox{2.2in}{!}{\includegraphics[scale=1]{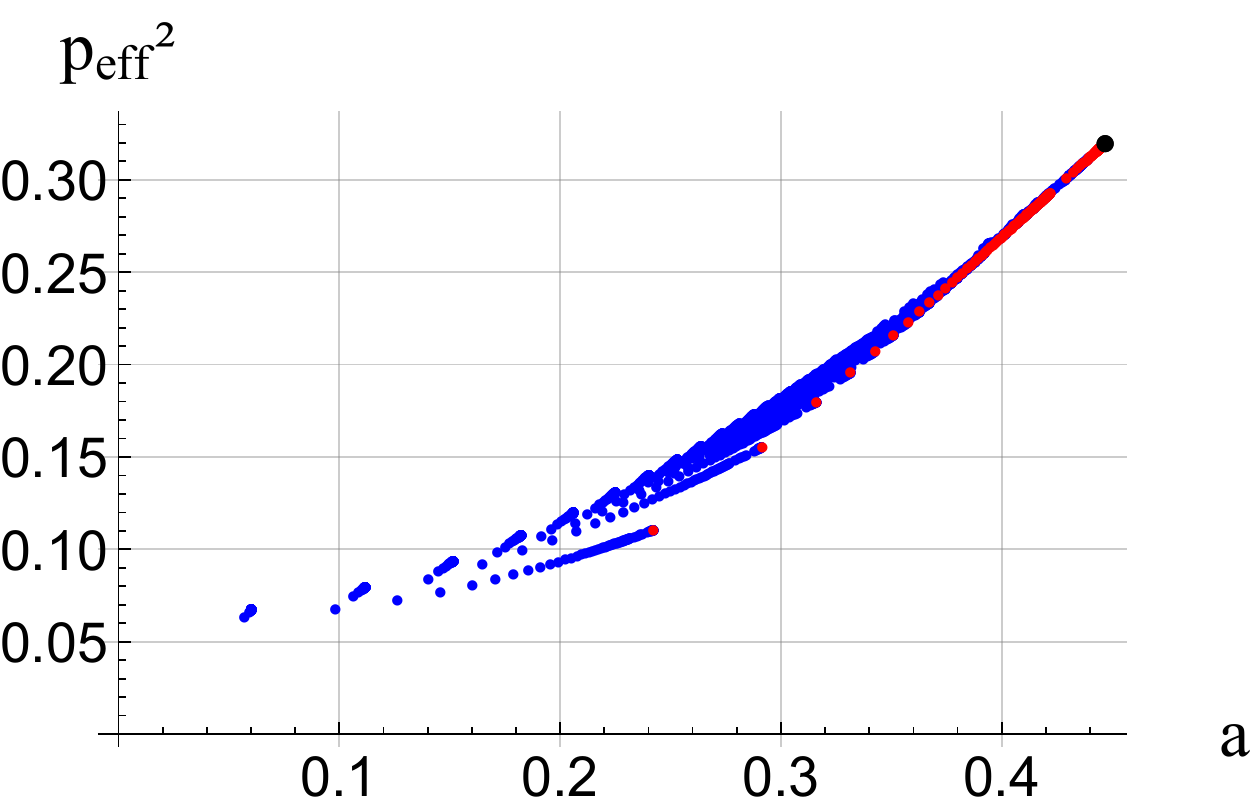}} \qquad
\resizebox{2.2in}{!}{\includegraphics[scale=1]{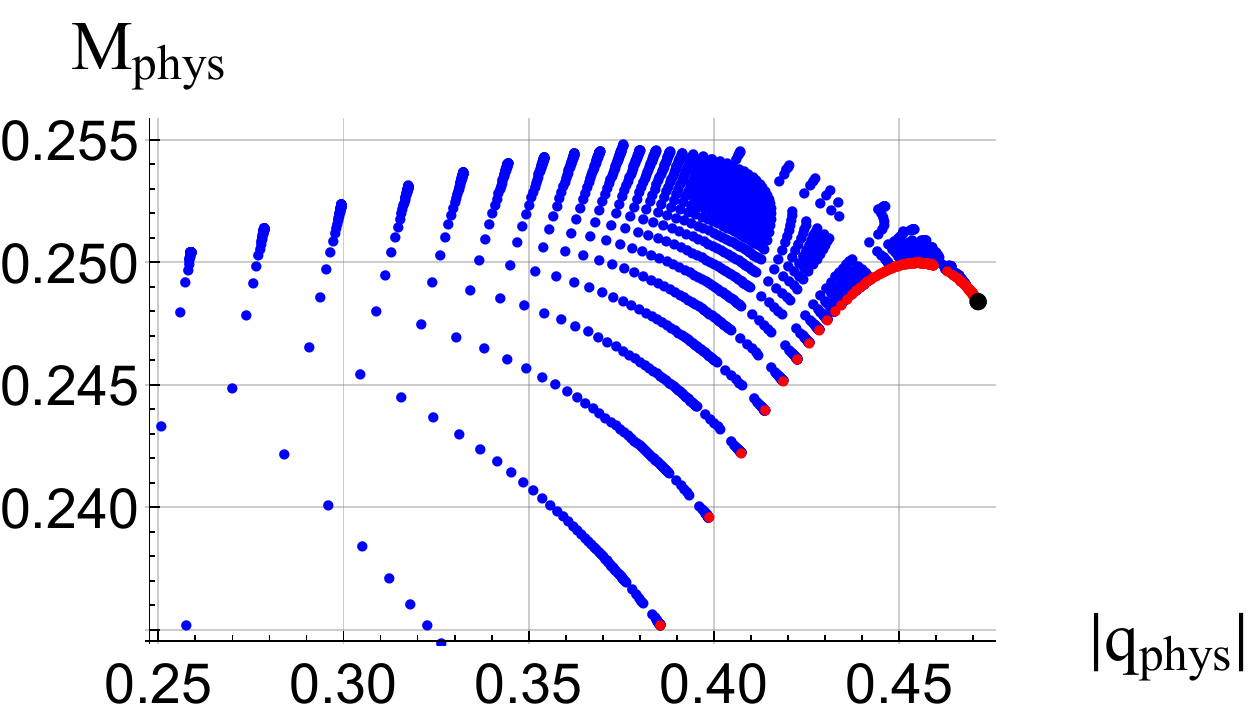}} \\
\resizebox{2.2in}{!}{\includegraphics[scale=1]{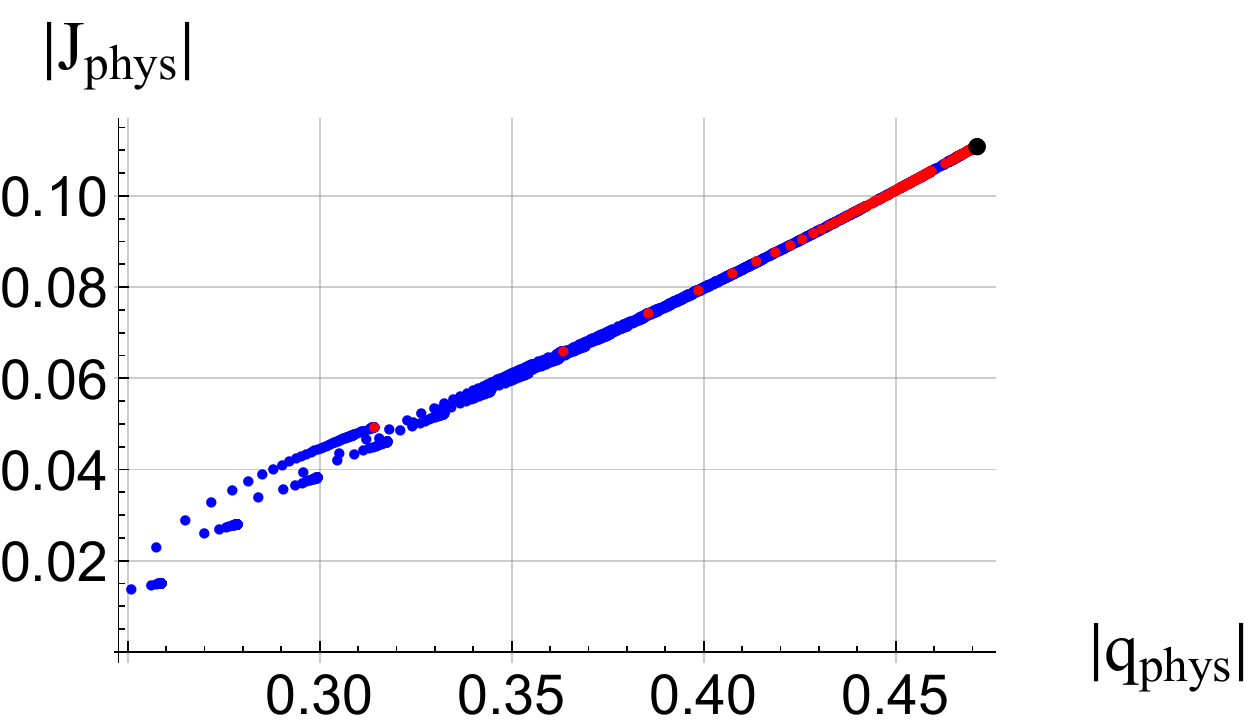}} \qquad
\resizebox{2.2in}{!}{\includegraphics[scale=1]{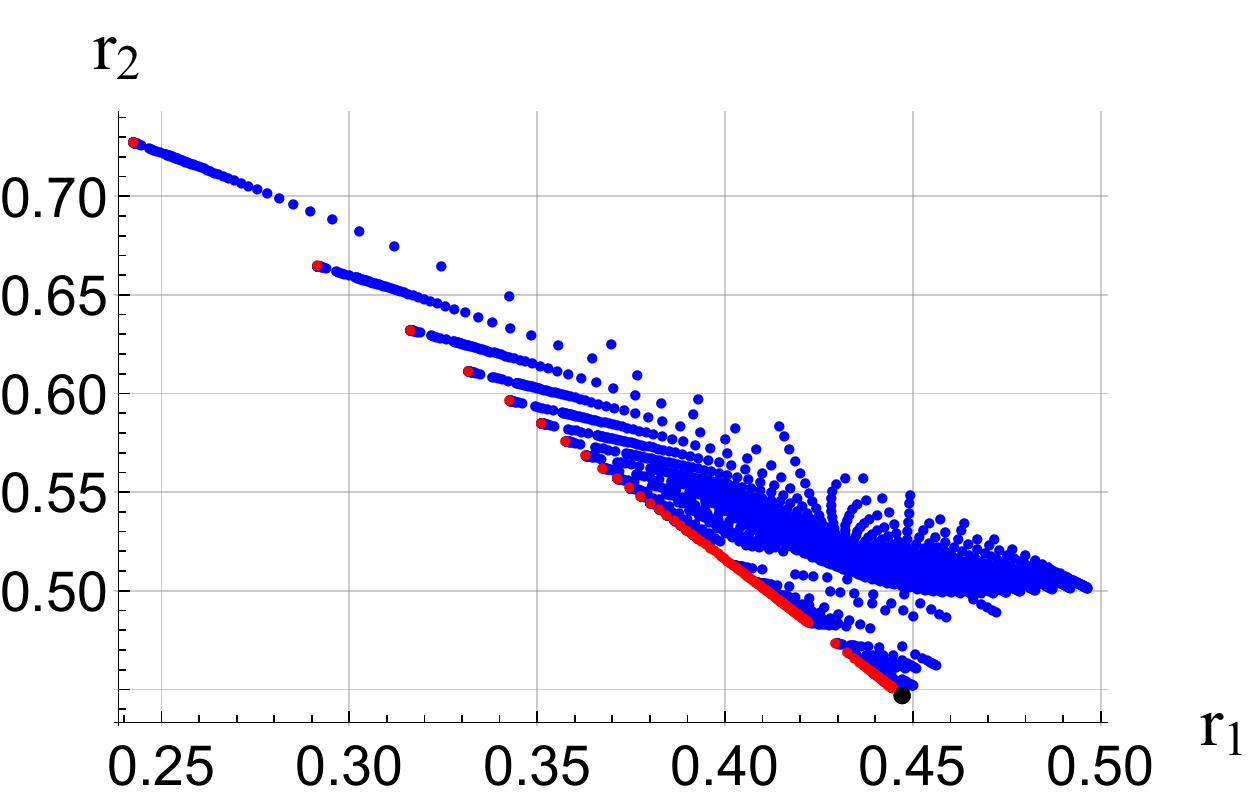}}
\caption{Correlations between $a$ and $\noQ$ (upper left plot),  $\ephysphys{}$  and $M_\phys$ (upper right), $\ephysphys$ and $|J_\phys|$ (lower left), and  $r_1$ vs.\ $r_2$ (lower right plot). The blue dots correspond to about 2000 solutions which are obtained by taking all values of $1\le n_2< n_1\le 50$ and a sample of values in the range $1\le n_2< n_1\le 1000$. The red dots are obtained by letting $n_1\to\infty$ (cf.~Section~\ref{s22VI15.1}), with $1\le n_2 \le 9900$. The black dot is the limit $n_1\to\infty$, $n_2\to \infty$ (cf.~Section~\ref{s22VI15.1}).}
\label{F21VI15.1}
\end{center}
\end{figure}

We expect that the threshold \eq{21VI15.31} is irrelevant, and indeed we have randomly sampled many further values of  $(n_1,n_2)$, including e.g.
$$
 (n_2,n_1)\in \{(1,10000),(20,1000), (200,1000),(1000,10000)\}
 \,,
$$
with the same result.
Plots displaying various correlations between parameters are shown in
Figure~\ref{F21VI15.1}.
The plots show that the resulting parameters $(a,M,\noQ)$ are bounded, and that  the values of the parameters approach affine correlations as both $n_1$ and $n_2$ tend to infinity.
This is explained in Section~\ref{s22VI15.1} below, where exact bounds and the asymptotically affine relations are derived.

%%%%%%%%%%%%%%%%%%%%%%%%%%%%%%%%%%%%%%%%%%%%%%%%%%%%%%%%%%%%%%%%%%%%%%%%%%%%%%%%%%%%%%%%%%%%%%%
%%%\input{physpar_corr_newCH}%%%%%%%%%%%%%%%%%%%%%%%%%%%%%%%%%%%%%%%%%%%%%%%%%%%%%%%%%%%%%%%%%%
\section{The limit $n_1 \to \infty$}
 \label{s22VI15.1}

An interesting case arises when we require $r =a$ to be a double zero of $\Delta_r$. While in this case the geometry is not compact anymore, the resulting manifold provides a  description of the geometry which is approached when $n_1$ tends to infinity with $n_2$ kept fixed. The values of the parameters $(a,m,\noQ )$ which arise in this case correspond to the limiting curves which arise in the plots showing the correlations between the parameters.

In order to study the system (\ref{15I12.0}-\ref{15I12.4}) for large $n_1$,  we rewrite \eq{15I12.4}  in the  form
\bena
%	\Delta_r(r_1)&=&0
%	\,, \label{15I12.0C}
%\\
%	\Delta_r(r_2)&=&0
%	\,, \label{15I12.1C}
%\\
%	\Delta'_r(r_1)   - \frac{2 a \Xi}{n_1} &=& 0
% 	\,, \label{15I12.2C}
%\\
%    -\Delta'_r(r_2)  n_2 - 2 a \Xi &=& 0
%	\,, \label{15I12.3C}
% \\
	(r_1^2-a^2)  -   (r_2^2-a^2) \frac{n_2}{n_1} &=&0
	\,. \label{15I12.4C}
\eena
Passing to the limit $n_1 \rightarrow \infty$ with $n_2$ fixed one is led to
\bena
 &
	0 = r_1^2-a^2 =
	\Delta'_r(r_1)    =
	\Delta_r(r_1) =
	\Delta_r(r_2)=
 	-\Delta'_r(r_2)  n_2 - 2 a \Xi
 	\,.
  &
  \label{15I12.3D}
\eena
In particular $r_1=a$. Scaling the metric by a constant so that $\Lambda=3$,
and using $r_1=a$ in Eq.(\ref{15I12.2})  we obtain
$M=a (1-a^2)$. Injecting in (\ref{15I12.0})
gives $\esquare{}=2a^2 (a^2-1)$.
Summarising
\bena
 &
r_1=a
  \,,
%  \label{15VI12.0}
%\\
\qquad
 M
  =a (1-a^2)
  \,,
 % \label{15VI12.1}
%\\
\qquad
 \esquare{}
  =2a^2 (1-a^2)
  \,.
 &
   \label{15VI12.2}
\eena
The parameter $a$ can then be determined using
\bena
	\Delta_r(r_2) = 0 =
	 \Delta'_r(r_2)  n_2 + 2 a \Xi
	\,, \label{15I12.3E}
\eena
with $M$ and $\esquare{}$ given by (\ref{15VI12.2}). Some algebra gives
\bena
 a
  &=&
   \sqrt{\frac{2 n_2 \left(5 n_2-2 \sqrt{8 n_2+1}+4\right)-\sqrt{8 n_2+1}+1}{2 		
   n_2 (25 n_2+8)+2}}
   \,, \label{15VI19.3}
\\
 r_2
  &=&\frac{\left(2 n_2+\sqrt{8 n_2+1}+1\right) \sqrt{\frac{4 n_2 \left(5 n_2-2
  \sqrt{8 n_2+1}+4\right)-2 \sqrt{8 n_2+1}+2}{n_2 (25 n_2+8)+1}}}{4 n_2}
   \,.
    \label{31XII15.1}
\eena

The corresponding physical parameters are
\bena
 M_{\phys} =\frac{a(1-a^2)}{(1+a^2)^2}
   \,,
   \qquad
 |J_{\phys}|
  = \frac{a^2(1-a^2)}{(1+a^2)^2}
   \,,
   \qquad
 \ephysphys
  = \frac{a\sqrt{2(1-a^2)}}{(1+a^2)}
   \,.
\eena
Here $M_{\phys} = M/(1+\lambda a^2)^2$ is the \emph{physical mass} {of the Lorentzian partner solution} (compare~\cite{GomberoffTeitelboim,CJKKerrdS}),   $|J_{\phys}| = a M/(1+\lambda a^2)$ is the \emph{Komar angular momentum} {of the Lorentzian partner solution}, and $\ephysphys := {\sqrt{\noQ}}/{(1 + \lambda a^2) }$
is the \emph{total magnetic Maxwell charge} of the Lorentzian solution with $e=0$ (compare~\cite{Sekiwa}).

We have
$$
 \Delta_r''|_{r=a} = 2 - 10 a^2
 \,,
$$
so that $\Delta_r $ is positive for $0<a<r_2$, with a simple zero at $r=r_2$, if and only if
\bel{20VI15.1}
 0 < a < \frac 1 {\sqrt 5}
 \,.
\ee

Inspection of   \eq{15III15.9} shows that the metric $g$ is complete, with a smooth axis of rotation at the other zero $r=r_2$ of $\Delta_r$ when $n_2\in \Z$. The set $r=a$ is infinitely far away, with the region $r\to a$ displaying an interesting geometry: While the circles of constant   $t$, $r$ and $\theta\not\in\{0,\pi\}$ shrink  to zero as $r$ tends to $a$,  the metric on the spheres of constant  $\varphi$ and $r$ is stretched along the meridians and approaches a smooth Riemannian metric on a cylinder obtained by removing the north and south pole from $S^2$.

\begin{figure}
  \begin{center}
    \includegraphics[width=.49\textwidth]{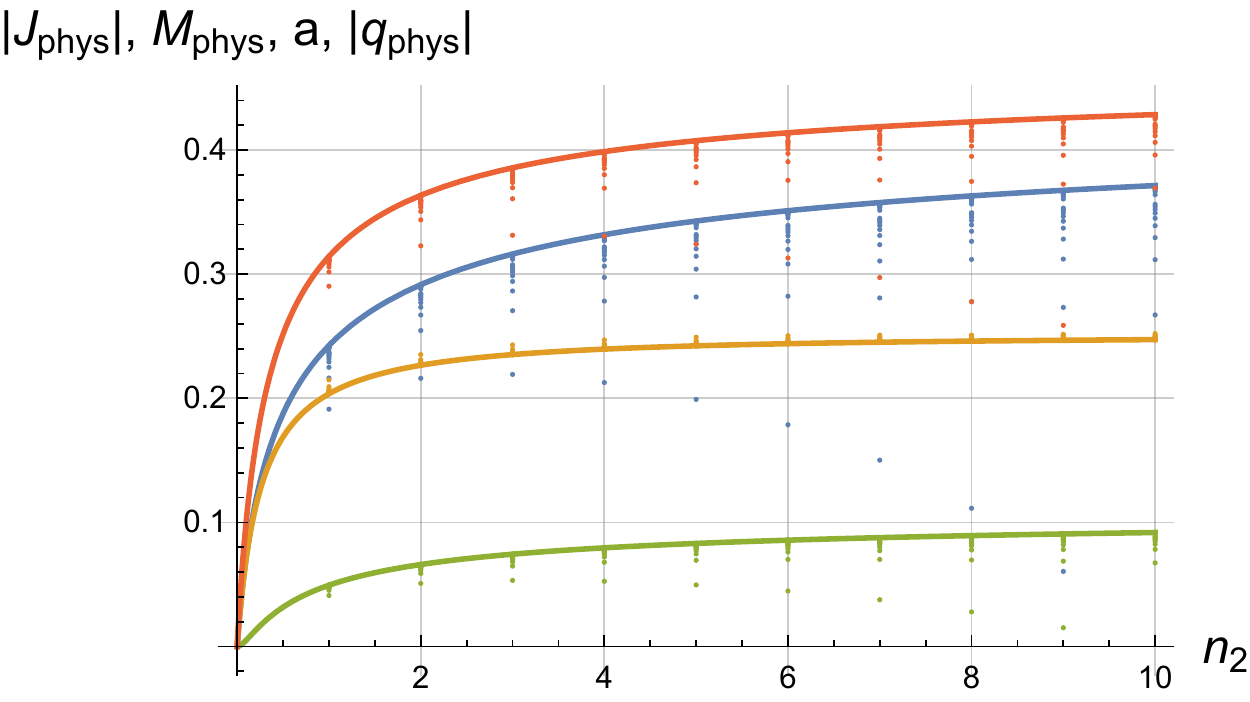}
    \
    \includegraphics[width=.49\textwidth]{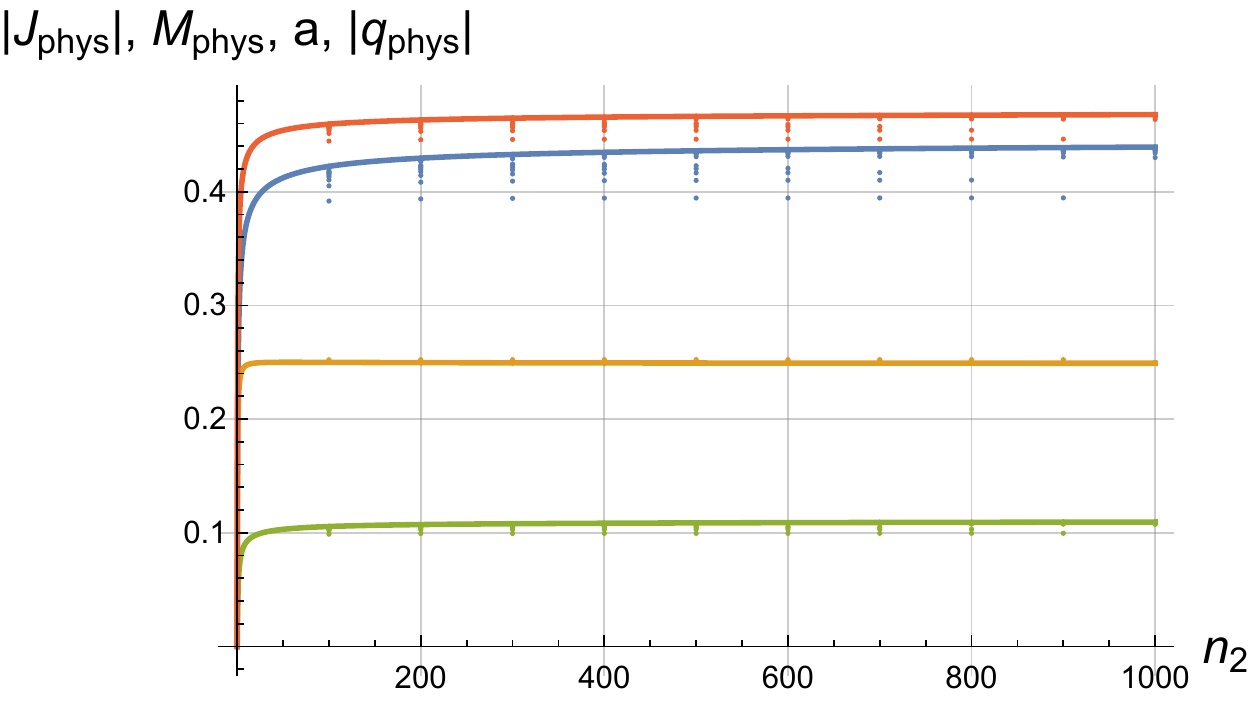}% MH plots with the dots on
  \end{center}
  \caption{\label{F20VI15.1} Plots of
  $|J_{\phys}|$ (lowest curve),
  $M_{\phys} $ (next to lowest on the left plot), $a$ (next to highest curve), and $\ephysphys$ (highest curve)
  as functions of a continuous variable $n_2\in[0,10]$ (left plot) and $n_2\in[0,1000]$ (right plot). The dots correspond to the values obtained for the solutions with the given values of $n_2$ and with $n_1$ increasing in logarithmic steps to $10000$  (left plot) and $100000$ (right plot).}
\end{figure}

\begin{figure}[h!]
  \begin{center}
    \includegraphics[width=.45\textwidth]{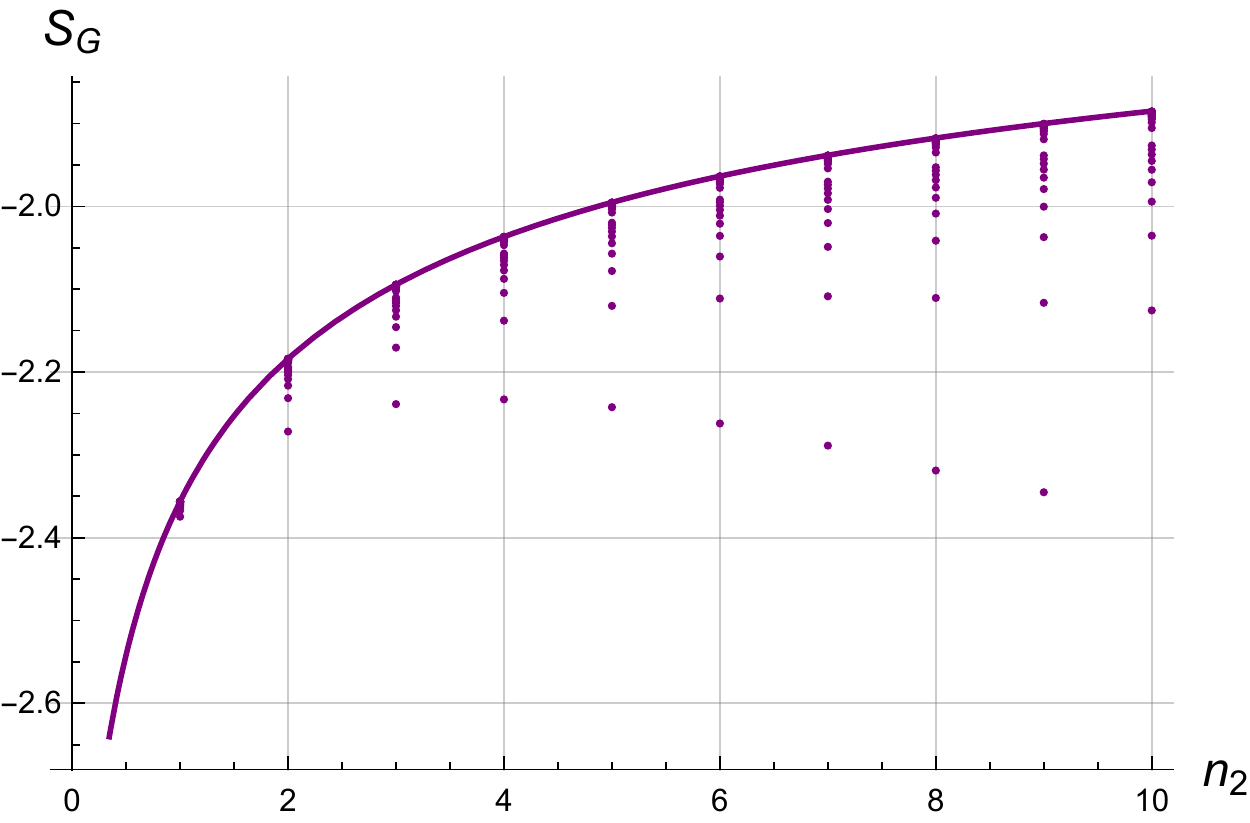}% MH plots with the dots on
    \quad
    \includegraphics[width=.45\textwidth]{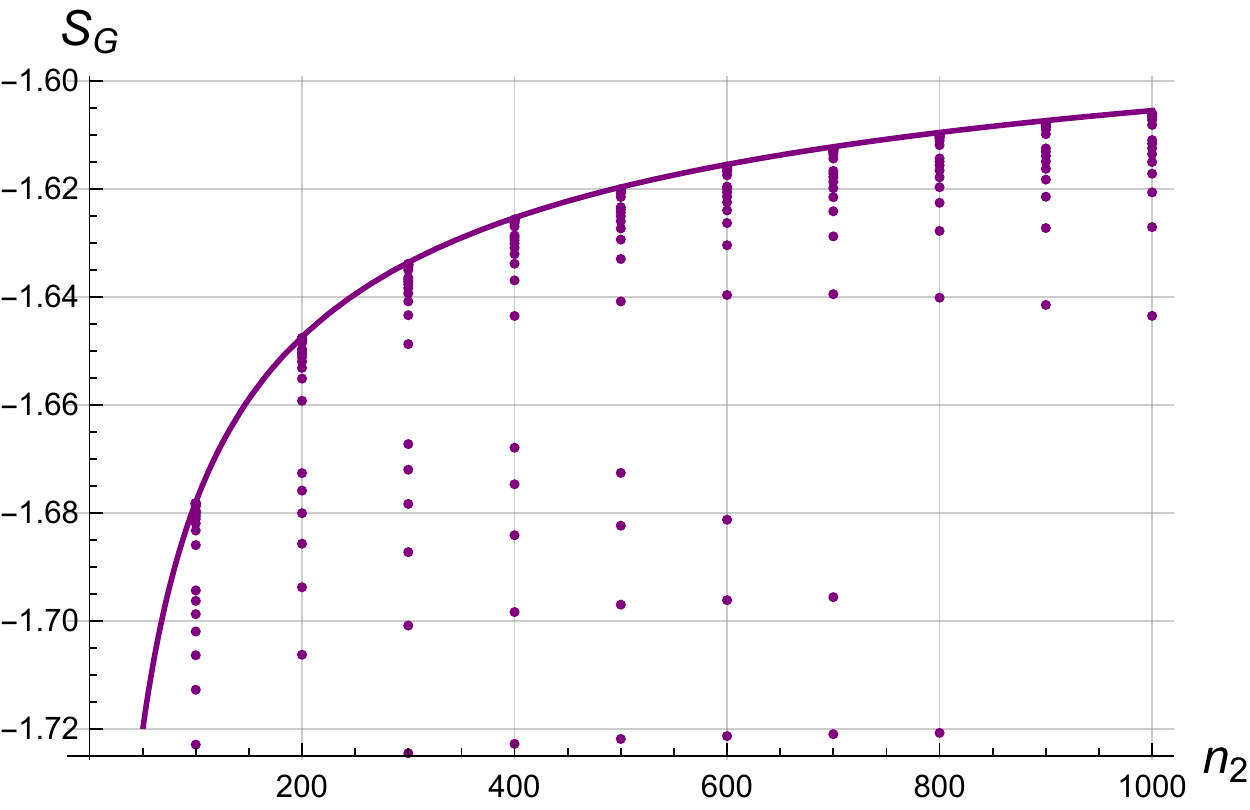}% MH plots with the dots on
  \end{center}
  \caption{\label{F20VI15.1a} Plots of the gravitational contribution $S_{G}=-\Lambda V/(8\pi )$, scaled to $\Lambda=3$, to the Euclidean action
  $S$ as function of a continuous variable $n_2\in[0,10]$ (left plot) and $n_2\in[0,1000]$ (right plot). The dots correspond to the values obtained for the solutions with the given values of $n_2$ and with $n_1$ increasing in logarithmic steps to $10000$  (left plot) and $100000$ (right plot).}
\end{figure}

We have the following expansions, for large  $n_2\in\N$,
\bena
 \nn
 a
  &=&
    \sqrt{\frac{2 n_2 \left(5 n_2-2 \sqrt{8 n_2+1}+4\right)-\sqrt{8 n_2+1}+1}{2 		n_2 (25 n_2+8)+2}}
\\
 \nn
 &=&
 \frac{1}{\sqrt{5}}-\frac{2}{5} \sqrt{\frac{2}{5}} \sqrt{\frac{1}{n_2}}+\frac{2}{25 \sqrt{5} n_2}
  +\frac{7}{100 \sqrt{10}}
   \left(\frac{1}{n_2}\right)^{3/2}
   %-\frac{11}{500 \sqrt{5} n_2^2}
   %-\frac{1623 \left(\frac{1}{n_2}\right)^{5/2}}{400000 \sqrt{10}}
   %+O\left(\left(\frac{1}{n_2}\right)^3\right)
   +O\left( {n_2} ^{-2}\right)
    \phantom{xxxx}
\\
 &&
  \to_{n_2\to \infty }    \frac 1{\sqrt 5} \approx 0.45
   \,,
\\
 \nn
 M
   &=& \sqrt{\frac{\left(2 n_2 \left(5 n_2-2 \sqrt{8 n_2+1}+4\right)-\sqrt{8
   n_2+1}+1\right) }{8(n_2 (25 n_2+8)+1)^3}} \times
\notag \\
 &&
    \phantom{xxxx}
      \times \left(4 n_2 \left(10 n_2+\sqrt{8
   n_2+1}+2\right)+\sqrt{8
   n_2+1}+1\right)
\notag \\
 &=&
 \frac{4}{5 \sqrt{5}}
  -\frac{4}{25} \sqrt{\frac{2}{5}} \sqrt{\frac{1}{n_2}}
   -\frac{4}{25 \sqrt{5} n_2}+\frac{39
   }{250 \sqrt{10}}   \left(\frac{1}{n_2}\right)^{3/2}
   %+\frac{7}{1250 \sqrt{5} n_2^2}-\frac{39831 %\left(\frac{1}{n_2}\right)^{5/2}}{1000000
   %\sqrt{10}}    +O\left(\left(\frac{1}{n_2}\right)^3\right)
   +O\left( {n_2} ^{-2}\right)
\notag \\
 &&
  \to_{n_2\to \infty }     \frac{4}{5 \sqrt{5}} \approx 0.36
   \,,
\\
 \noQ
   &=&\frac{n_2^2 \left(4 n_2 \left(50 n_2-15 \sqrt{8 n_2+1}+34\right)-15
   \sqrt{8 n_2+1}+17\right)}{(n_2 (25 n_2+8)+1)^2}
\notag \\
 &=&\frac{1}{\sqrt{5}}-\frac{2}{5} \sqrt{\frac{2}{5}} \sqrt{\frac{1}{n_2}}
 +\frac{2}{25 \sqrt{5} n_2}
  +\frac{7 }{100 \sqrt{10}} \left(\frac{1}{n_2}\right)^{3/2}
   %-\frac{11}{500 \sqrt{5} n_2^2}
   %-\frac{1623 \left(\frac{1}{n_2}\right)^{5/2}}{400000 \sqrt{10}}
   %+O\left(\left(\frac{1}{n_2}\right)^3\right)
   +O\left( {n_2} ^{-2}\right)
\notag \\
 &&
  \to_{n_2\to \infty }    \frac 8 {25} \approx 0.32
   \,,
    \label{30XII15.1}
\eena
\bena
 M_{\phys}
  &=&\frac{4 n_2 \left(10 n_2+\sqrt{8 n_2+1}+2\right)+\sqrt{8 n_2+1}+1}{\left(4
  n_2 \left(-15 n_2+\sqrt{8 n_2+1}-6\right)+\sqrt{8 n_2+1}-3\right)^2}
	\notag 
\\
  &\phantom{=}&
   \times \sqrt{2 (n_2 (25 n_2+8)+1) \left(2 n_2 \left(5 n_2-2 \sqrt{8
  n_2+1}+4\right)-\sqrt{8 n_2+1}+1\right)}
 \phantom{xxxxx}
   \label{29VI15.2}
  \\
 &=&
  \frac{\sqrt{5}}{9}+\frac{1}{27} \sqrt{\frac{2}{5}} \sqrt{\frac{1}{n_2}}-\frac{1}	
  {5 \sqrt{5} n_2}+
  \frac{421 \left(\frac{1}{n_2}\right)^{3/2}}{48600 \sqrt{10}}+\frac{8867}{145800 
  \sqrt{5} n_2^2}-
  O\left( {n_2} ^{-5/2}\right) 
\notag 
\\
 &&
  \to_{n_2\to \infty }  \frac{  \sqrt{5}}{9}  \approx 0.25
   \,,
\\
 |J_{\phys}|
   &=&\frac{8 n_2^2}{4 n_2 \left(18 n_2+3 \sqrt{8 n_2+1}+10\right)+3 \sqrt{8 n_2+1}+5}
   \label{29VI15.3}
\\
 &=&
 \frac{1}{9}-\frac{1}{27} \sqrt{2} \sqrt{\frac{1}{n_2}}-\frac{1}{27 n_2}+\frac{83 }{1944
   \sqrt{2}}\left(\frac{1}{n_2}\right)^{3/2}+\frac{37}{5832 n_2^2}+O\left(n_2^{-5/2}\right)
\notag \\
 &  &
 \to_{n_2 \to \infty}
 \frac 19
   \,,
\\
 \ephysphys{}
   &=&\frac{2 (n_2 (25 n_2+8)+1)}{4 n_2 \left(15 n_2-\sqrt{8 n_2+1}+6\right)-\sqrt{8 n_2+1}+3}
\notag \\
  &\phantom{=}&
\times \sqrt{\frac{n_2^2 \left(4 n_2 \left(50 n_2-15 \sqrt{8 n_2+1}+34\right)-15 \sqrt{8 n_2+1}+17\right)}{(n_2 (25 n_2+8)+1)^2}}
   \label{29VI15.3a}
\\
 &=&
\frac{\sqrt{2}}{3}-\frac{1}{9}\sqrt{\frac{1}{n_2}}-\frac{7}{54 \sqrt{2} n_2}+\frac{55 }{1296}\left(\frac{1}{n_2}\right)^{3/2}+\frac{5}{243 \sqrt{2} n_2^2}+O\left(n_2^{-5/2}\right)
\notag \\
 &  &
 \to_{n_2 \to \infty}
 \frac{\sqrt{2}}{3} \approx 0.471
   \,.
\eea
Perhaps surprisingly, the total volume
of the solutions
(directly related to the gravitational contribution $S_{G}$ to the action, see \eq{30XII15.2} below) turns out to be  finite. To determine it we use \eq{31XII15.2} below with $\kappa=|\kappa_2|$, which   equals
\bean
 \kappa_2
  & = &
   \frac{-2 r_2
   \left(r_2^2-a^2\right)+
   2 a \left(a^2-1\right)+2
   r_2
   \left(1-r_2^2\right)}{2
   \left(1-a^2\right)
   \left(r_2^2-a^2\right)}
\\
 & = &
 -\frac{1}{2} \sqrt{\frac{5}{2}}
   \sqrt{\frac{1}{n_2}}
    +\frac{7}{8 \sqrt{5}
   n_2}
%   -\frac{81
%   \left(\frac{1}{n_2}\right)^{3/2}}{160
%   \sqrt{10}}+\frac{51}{400
%   \sqrt{5}
%   n_2^2}+O\left(\left(\frac{1}{n_2}\right)^{3/2} \right)
    +O\left(n_2^{-3/2}\right)
 \,.
\eeal{31XII15.3}
One finds
\bean
   V
   &=& \frac{ \pi^2  \left(4 n_2+\sqrt{8 n_2+1}-1\right)}{3 n_2}
   \label{29VI15.3b}
\\
 &=&
  \frac{4\pi^2 }{3}
  + \frac{4\pi^2}{3 \sqrt{2}} \sqrt{\frac{1}{n_2}}
   - \frac{\pi^2 }{3 n_2}
%  + \frac{\pi^2  }{12 \sqrt{2}} \left(\frac{1}{n_2}\right)^{3/2}
+O\left(n_2^{-3/2}\right)
\notag \\
 &  &
 \to_{n_2 \to \infty}
 \frac{4\pi^2}{3} \approx  13.16
   \,.
    \label{8IX15.6}
\eena
Plots showing monotonicity of some of the functions above, at least for $n_2$ large enough, can be found in Figure~\ref{F20VI15.1}. A plot of $S_{G}$ as a function of $n_2$ can be found in Figure~\ref{F20VI15.1a}.

$M_\phys$ attains its maximum at $n_2=\sqrt{\frac{1}{2} \left(799+565 \sqrt{2}\right)}+10 \sqrt{2}+14\approx 56.409$, at which point it equals $1/4$. Closer inspection, taking into account that we are only interested in integer values of $n_2$, gives
\bel{29VI15.5}
0.20361015\approx M_\phys|_{n_2=1} \le  M_\phys  \le M_\phys|_{n_2=56}\approx 0.24999998
 \,,
\ee
with the bounds being optimal.

All quantities have an asymptotic expansion, as $n_2$ tends to infinity, in terms of negative powers of $\sqrt{n_2}$.
This leads to simple relations between various quantities for $n_1$ and $n_2$ large, as follows: for large $n_2$ we have the approximate relations
\bena
  \sqrt{\frac{1}{n_2}}
 & \approx &
 \frac{5}{2} \sqrt{\frac{5}{2}}  \left( \frac{1}{\sqrt{5}} -a\right)
  \approx
 \frac{25}{4} \sqrt{\frac{5}{2}}
 \left( \frac{4}{5 \sqrt{5}} - M \right)
  \approx
 \frac{5}{2} \sqrt{\frac{5}{2}} \left(\frac{1}{\sqrt{5}} - \noQ \right)
 \approx
  \frac{2 \sqrt{2}}{\pi} \left( - \frac{\pi}{2} - S_{G} \right)
 \nn
\\
  &\approx &
27 \sqrt{\frac{5}{2}}  \left(   M_{\phys} - \frac{\sqrt{5}}{9} \right)
 \approx
  \frac{27}{\sqrt{2}}\left( \frac{1}{9}-|J_{\phys}|\right)
  \approx
  9 \left( - \frac{\sqrt{3}}{2} - \ephysphys{} \right)
   \,.
\eena
From this one obtains various approximately affine relations between the quantities above for $1 \ll n_1 \ll n_2$, e.g.
\beal{21VI15.41}
 |J_\phys|
  & \approx & -\sqrt 5
  M_\phys + \frac{2}{3}
   \,,
\\
 \ephysphys
  & \approx & -\frac{9}{3} M_\phys +\frac{\sqrt{2} -\sqrt{5}}{3}
   \,,
\\
 S_{G}
  & \approx & -\frac{\sqrt{5} \pi}{4}
  M_\phys -\frac{13}{36} \pi
   \,.
\eeal{21VI15.42}

\begin{figure}[h!]
  \begin{center}
     \includegraphics[width=.45\textwidth]{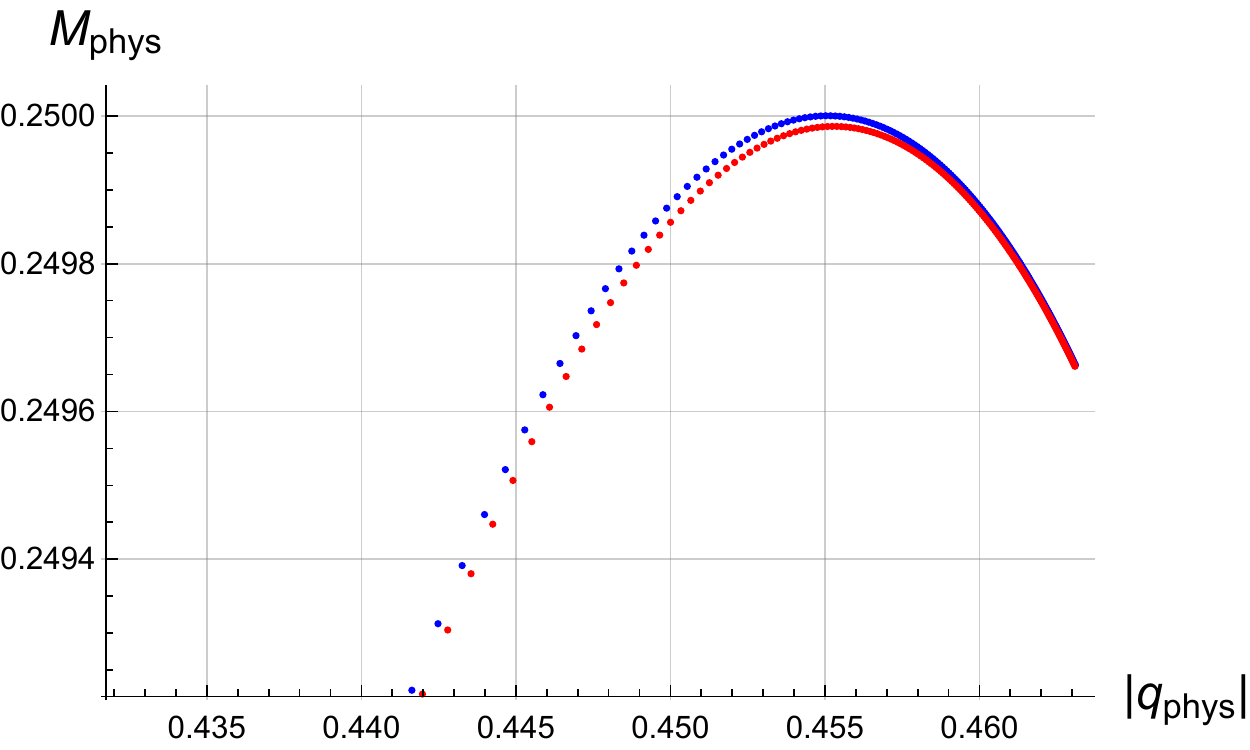}% MH plots with the dots on
  \end{center}
  \caption{\label{F29VI15.1} Correlation plot  in the
  $(|q_{\phys}|,M_{\phys})$ plane in the limit $n_1\to\infty$. The red points lie on the curve \eq{29VI15.1},  the blue dots arise from the exact solutions \eq{29VI15.2} and \eq{29VI15.3}.}
\end{figure}

One can similarly make a second-order approximation in $1/n$, by expanding the quantities of interest up to $o(n^{-1})$ and eliminating $n$ from the equations. As an example, near the maximum value of $|q_\phys|$ we obtain the relation 
\bel{29VI15.1}
M_\phys\approx \frac {3402 \sqrt{2} | 
   q_\phys | +394\sqrt {3}\sqrt {5 - 7\sqrt {2} | q_\phys | } - 
    1437} {2205\sqrt {5}} 
\,.
\ee
The exact solution  and the curve resulting from the second order approximation in $1/n$ can be seen in Figure~\ref{F29VI15.1}.

\begin{figure}[h!]
  \begin{center}
    \includegraphics[width=.4 \textwidth]{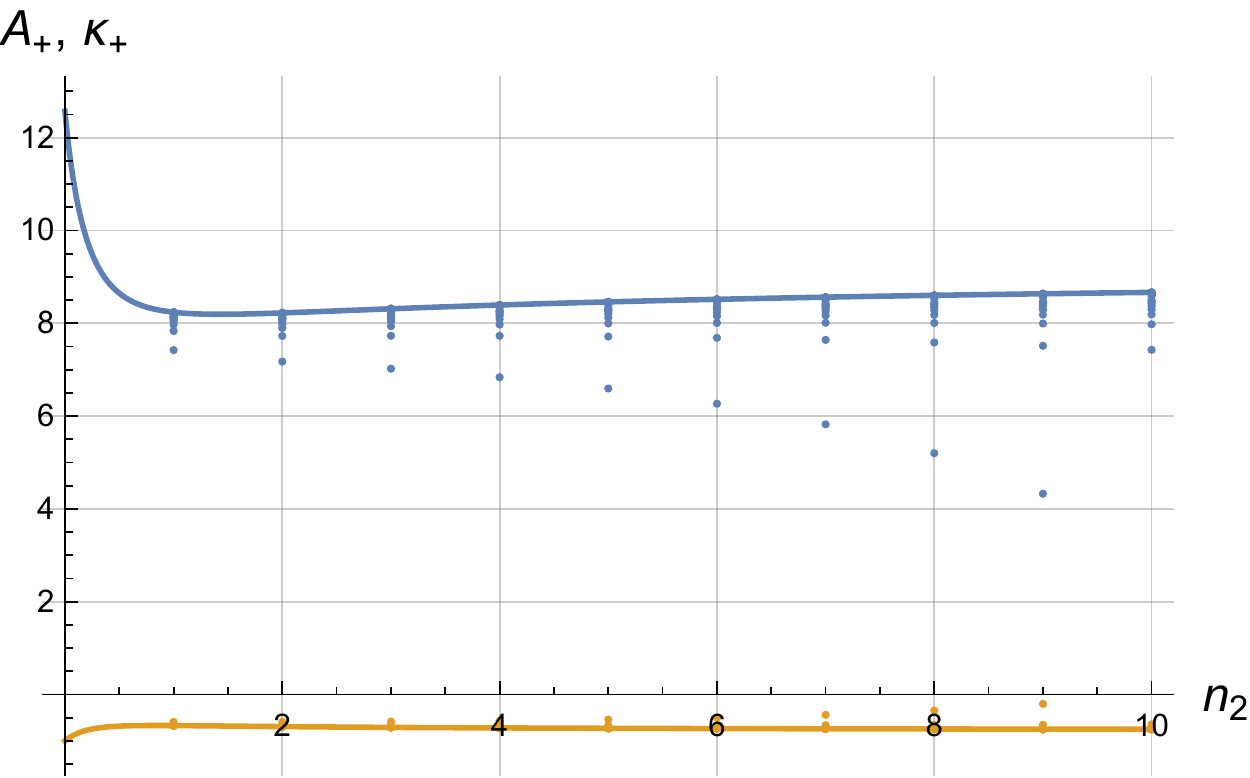}% MH plots with the dots on
    \qquad
    \includegraphics[width=.4 \textwidth]{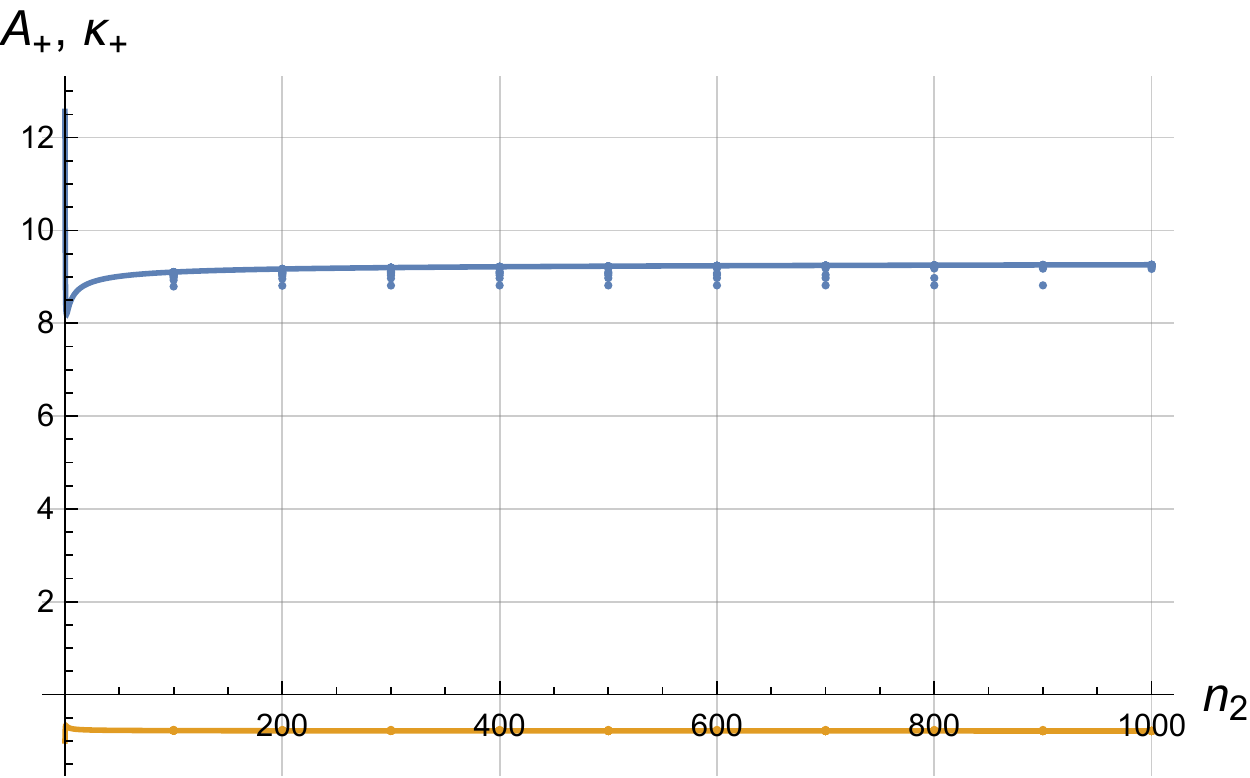}% MH plots with the dots on
  \end{center}
  \caption{\label{F20VI15.1b} Plots of
  $A_+$(blue line) and $\kappa_+$(orange line) as functions of a continuous variable $n_2\in[0,10]$ (left plot) and $n_2\in[0,1000]$ (right plot). The dots correspond to the values obtained for the solutions with the given values of $n_2$ and with $n_1$ increasing in logarithmic steps to $10000$  (left plot) and $100000$ (right plot).}
\end{figure}

In Figure~\ref{F20VI15.1b} we  plot the dependence on the continuous variable $n_2$, in the $n_1 \to \infty$ limit, of the area of the cross section of the horizon $A_+$ and the surface gravity $\kappa_+$ in the partner Lorentzian solutions.

In  Appendix~\ref{SI} the reader will find a translation of some of the numerical values above to SI units.
%%%%%%%%%%%%%%%%%%%%%%%%%%%%%%%%%%%%%%%%%%%%%%%%%%%%%%%%%%%%%%%%%%%%%%%%%%%%%%%%%%%%%%%%%%%%%%%

\section{Dirac strings}
 \label{s1XII15.1}

Similarly to~\cite{DuffMadore}, the existence of charged spinor fields on the Euclidean manifold leads to further constraints on the parameters of the solution. Indeed, comparing \eq{1I16.4} with \eq{1I16.4+} shows that the transition from a gauge potential  which is regular near $\{\cos(\theta)=1\,,\  r=r_i\}$ to a gauge potential  which is regular near $\{\cos(\theta)=-1\,,\  r=r_i\}$ requires a gauge transformation
$$
 A\mapsto A +  \frac{ 2p  }{\Xi}  ( \dnotdelta \varphi
 - \frac{a  }{ a^2-r_i^2 } \dnotdelta t)
   \,.
$$
If a Dirac field $\psi$ carries a charge $q_0$, such a gauge transformation induces a transformation
$$
 \psi \mapsto \exp\left(\frac{2 i q_0   p}{\hbar \Xi }\left(\varphi  - \frac{a  }{ a^2-r_i^2 } t\right)\right) \psi
   \,.
$$
Recall that $\varphi$ is $2\pi/k$-periodic, with $k=1$ except in the Page limit where $k=2$ can arise and which needs to be analysed separately in any case, see Section~\ref{ss22I16.1} below. Thus in the remainder of this section we assume that $\varphi$ is $2\pi$ periodic.
The requirement of single-valuedness of $\psi$ results in the condition
\bel{1XII15.13}
 \frac{2   q_0   p}{ \hbar \Xi} =: \hat n_1 \in \Z
  \,.
\ee

Next, \eq{1I16.4} and \eq{1XII15.1} show   that the gauge potentials
\bel{1I16.11}
  A + \frac{ p  }{\Xi}  ( \dnotdelta \varphi
 - \frac{a  }{ a^2-r_i^2 } \dnotdelta t)
 - \frac{e r_i }{ \Xi (a^2-r_i^2) } \dnotdelta t
 \,,
 \quad
 i=1,2,
\ee
are regular near  $r=r_i$ and $\theta=0$. Passing from one to the other requires a gauge transformation
$$
 A\mapsto A+ \left(\frac{pa+ e r_2 }{ \Xi (a^2-r_2^2) }
 - \frac{pa+e r_1 }{ \Xi (a^2-r_1^2) }
 \right)
  \dnotdelta t
  \,.
$$
Keeping in mind that $ t $ has period $ 2\pi \omega$, the associated transformation of the spinor field $\psi$ leads to the further condition
\bel{1XII15.14}
 \underbrace{\frac{    q_0  }{\hbar \Xi}\bigg(    p}_{\hat n_1/2}     \big(
 \underbrace{\frac{ a \omega }{    r_1^2-a^2  }}_{n_1}  -
  \underbrace{\frac{ a \omega  }{    r_2^2-a^2  }
                                                    }_{n_2}
        \big)
        +
  e \omega   \big(\frac{  r_1 }{    r_1^2-a^2  }  - \frac{  r_2 }{    r_2^2-a^2  }
        \big)
       \bigg)
            \in \Z^*
  \,.
\ee
A similar analysis near $\theta=\pi$ leads to the further condition
\bel{1XII15.14+}
 \underbrace{\frac{    q_0  }{\hbar \Xi}\bigg(    p}_{\hat n_1/2}     \big(
 \underbrace{\frac{ a \omega }{    r_1^2-a^2  }}_{n_1}  -
  \underbrace{\frac{ a \omega  }{    r_2^2-a^2  }
                                                    }_{n_2}
        \big)
       -
  e \omega   \big(\frac{  r_1 }{    r_1^2-a^2  }  - \frac{  r_2 }{    r_2^2-a^2  }
        \big)
       \bigg)
            \in \Z^*
  \,.
\ee
We conclude that we must have
\bel{1XII15.14++}
 \frac{\hat n_1 ({n_1}  - {n_2})}{2}\in \N^*
  \,,
\ee
and that there exists $\hat n_2 \in \N^*$ so that
\bel{1XII15.14--}
  e \omega   \left(\frac{  r_1 }{    r_1^2-a^2  }  - \frac{  r_2 }{    r_2^2-a^2  }
        \right)
       = \hat n_2
  \,.
\ee
Eliminating $q_0$ between \eq{1XII15.13} and \eq{1XII15.14} imposes a quantised relation between $p$ and $e$:
\bel{1XII15.15}
 p =
  \underbrace{ \frac{    \omega }{ 2   }
  \left(\frac{  r_1 }{    r_1^2-a^2  }  - \frac{  r_2 }{    r_2^2-a^2  }
        \right)
        }_{=:\sigma}
         \times  \frac{    \hat n_1   }{   \hat n_2 } \times
  e
  \,.
\ee

Recall that given a set $(M,a,\noQ)$, parameterised by two integers $(n_1,n_2)$ with $n_1<n_2$  and arising from a smooth compact Riemannian solution, we have so far been associating to it a Lorentzian partner solution with the same values of $M$ and $a$, with $p^2=\noQ$ and with $e=0$. However, if one adds the requirement of well-defined charges spinor fields to the picture, instead of choosing $e=0$ on the Lorentzian side one might wish to request that \eq{1XII15.15} holds. This adds two further quantum numbers $(\hat n_1, \hat n_2)$ to the picture.  Taking into account the inequality $\noQ= p^2 -e^2>0$
one is led to the condition
\bel{2XII15.1}
 \lnn > 1
 \,.
\ee
Given a pair $(\hat n_1,\hat n_2)$ such that \eq{2XII15.1} holds (note that this can always be achieved by choosing $\hat n_1$ large enough), we can determine $|q_0|$, $|e|$ and $|p|$ from \eq{1XII15.13}-\eq{1XII15.15}:
\bean
 &
  \displaystyle
 |e| =
  \sqrt{\frac  {\noQ }{(\lnn)^2-1}}
 \,,
  \quad
 |p| = {\lnn}\sqrt{\frac  {\noQ }{(\lnn)^2-1}}
 \,,
% &
%\\
% &
 \quad
 |q_0|
  =
   \frac {\hbar \Xi} 2  \sqrt{\frac{\sigma^2 \hat n_1^2-\hat n_2^2} {\noQ }}
 \,.
 &
\eeal{2XII15.2asdf}
In this way we are led to a discrete family of solutions parameterised by four integers $(n_1,n_2,\hat n_1, \hat n_2)$ subject to the constraints   \eq{1XII15.14++} and \eq{2XII15.1}.
%, with a unique minimal value of Dirac-test-field-charge $|q_0|$ associated with %each such solution obtained when $(\hat n_1, \hat n_2)$ are relatively prime.

It holds that $\noQ < p^2\to_{(\hat n_1/\hat n_2) \to \infty} {\noQ}$, $e\to_{(\hat n_1/\hat n_2) \to \infty} 0$, and thus $\noQ < p^2+e^2 \to_{(\hat n_1/\hat n_2) \to \infty} {\noQ}$.

The global structure of the resulting Lorentzian partners is the same as in the case $e=0$, see Figure~\ref{F25VI12.7}.

\appendix
%%%\input{exampleCH}
%%%%%%%%%%%%%%%%%%%%%%%%%%%%%%%%%%%%%%%%%%%%%%%%%%%%%%%%%%%%%%%%%%%%%%%%%%%%%%%%%%%%%%%%%%%%%%%

\section{A typical solution}
 \label{A21VI15.1}

We rescale the metric so that $\lambda=1$. We choose $n_1 = 10$, $n_2 = 9$.
  With this choice the system (\ref{15I12.0}-\ref{15I12.4}) takes the explicit form
\bena
-a^2+\noQ -2 M r_1+r_1^2+a^2 r_1^2-r_1^4&=&0 \,, \nonumber \\
%-a^2 + \noQ  - 2 M r_2 + r_2^2 + a^2 r_2^2 - r_2^4&=&0 \,, \nonumber \\
-2 a + 2 a^3 - 20 M + 20 r_1 + 20 a^2 r_1 - 40 r_1^3&=&0  \,, \nonumber \\
-2 a + 2 a^3 + 18 M - 18 r_2 - 18 a^2 r_2 + 36 r_2^3&=&0 \,, \nonumber \\
10 (r_1^2-a^2) -9 (r_2^2 -a^2)&=&0 \,,
\label{19V15.4}
\eena
as well as an equation for $r_2$ identical to the first  equation  above.
The Buchberger algorithm for finding a Gr\"obner basis for Eq.(\ref{19V15.4}), as implemented in {\sc Mathematica}, yields the following
system
\begin{tiny}
\bena
141447860388864000000 (\noQ) ^2-2530102285619187840000 (\noQ) ^3+6902836371659336516100 (\noQ) ^4 &\phantom{=}& \nonumber \\
-7443462023036715884580 (\noQ) ^5+3324944139689702617201 (\noQ) ^6&=&0\,, \nonumber \\ \nonumber \\
269121969463191443505728626849880910201000 (\noQ) ^2+5123491133454465890342571180870383758599000 a^2 (\noQ) ^2  &\phantom{=}& \nonumber \\
-4306620505226193997812468562360852027723500 (\noQ) ^3+661016788713267074222610725116146960042870 (\noQ) ^4  &\phantom{=}& \nonumber \\
-1418820167927814403912453122762613275222257 (\noQ) ^5 &=&0\,,  \nonumber \\ \nonumber \\
1840946142733449839332390348051522429788563882649600000 M (\noQ)  &\phantom{=}& \nonumber \\
-140403030498229867043777134104718536116027658797039040000 a (\noQ) ^2 &\phantom{=}& \nonumber \\
+425794621585557982844978758649217892223137640430823652700 a (\noQ) ^3  &\phantom{=}& \nonumber \\
-476718734408676529956326149018521215355879124052827578300 a (\noQ) ^4  &\phantom{=}& \nonumber \\
+216318197798255246294998226248424687617679013902862944743 a (\noQ) ^5 &=&0\,,  \nonumber \\ \nonumber \\
10604338062917514381295873956265388153861661532099121130946560000000 a^3 &\phantom{=}& \nonumber \\
-10604338062917514381295873956265388153861661532099121130946560000000 a^5 &\phantom{=}& \nonumber \\
+1767389677152919063549312326044231358976943588683186855157760000000 a (\noQ)  &\phantom{=}& \nonumber \\
-588227304146277718291223364304260174172128286418763128059980799680000 a (\noQ) ^2 &\phantom{=}& \nonumber \\
+1755642879359412179543337165217487045071628171531789605489627506273900 a (\noQ) ^3 &\phantom{=}& \nonumber \\
-1954502602810413009680346541725810677017518746391376355441883692255820 a (\noQ) ^4 &\phantom{=}& \nonumber \\
+893243009923318159877420930128536388883746991239505692176977591346539 a (\noQ) ^5 &\phantom{=}& \nonumber \\
+3534779354305838127098624652088462717953887177366373710315520000000 (\noQ)  r_2 &=&0\,,  \nonumber \\
\label{19V15.5}
\eena
\end{tiny}
together with an identical equation for $r_1$.

The structure of the equations is typical in the following sense: Since {\sc Mathematica} does not manage to find a Gr\"obner basis when $n_1$ and $n_2$ are left as general parameters, our procedure is to provide the values of $n_1$ and $n_2$ and then seek the basis. All the resulting polynomials that we have inspected have then a structure identical to the one above.

It can be seen that solving the system (\ref{19V15.5}) in the manner  described above requires only solving polynomial equations in a single variable
of at most forth order, and so explicit analytic expressions can be given. However, the expressions obtained, especially for $r_1$ and $r_2$, become very unwieldy. Therefore,  instead
of the full analytic expressions, we give
only the first five nontrivial digits after the decimal point of the parameters for the solution of (\ref{19V15.4}) that fulfills the constraints:
\be
 r_1=0.48613 \,,~~r_2=0.51203 \,,~~M=0.25211 \,,~~a=0.060481 \,,~~\noQ =0.067439
  \,.
\ee

\section{Physical quantities}
 \label{s31XII15.4}

\subsection{Euclidean case}

The ``surface gravity'' of the zeros of the $\partial_t$-Killing vector, located at $r_1$ and $r_2$, reads
\bena
	\kappa:= \frac{1}{2 \Xi (r_i^2-a^2)} \Delta_{r} ' |_{r={r_1}}
=-\frac{1}{2 \Xi (r_i^2-a^2)} \Delta_{r} ' |_{r={r_2}}
 \,.
\eena

Since $\partial_\varphi$ and $\partial_t$ are Killing fields
and
$$
 (t,r,\theta,\varphi) \in [0, \frac{2 \pi}{\kappa}) \times [r_1,r_2) \times [0,\pi)  \times [0 , 2 \pi)
 \,,
$$
we obtain the following formula for the areas of the zero-set of $\partial_t$, located at $r_1$ and $r_2$,
\bena
	A_i&=& 2 \pi \int_0^\pi \sqrt{g_{\varphi \varphi} g_{\theta \theta}} |_{r={r_i}} \,
	d \theta
   = \pi \int_0^\pi\frac{\left(r_i^2-a^2\right) \sin (\theta )}{\Xi } \, d \theta
	\, \nonumber
\\
    &=&   \frac{4 \pi  \left(r_i^2-a^2\right)}{\Xi }
    \,
\eena
and for the volume of the manifold
\bena
	V&=& 2 \pi \frac{2 \pi}{\kappa} \int_{r_1}^{r_2} \int_0^\pi \sqrt{g} \, d \theta d r
	 = \frac{4 \pi^2}{\kappa} \int_{r_1}^{r_2} \int_0^\pi \frac{\Sigma  \sin (\theta )}{\Xi 	^2} \, d \theta d r
	\, \nonumber
\\
	&=& \frac{8 \pi ^2}{{3 \kappa  \Xi ^2}} \left[\left(r_2^3-r_1^3\right)-a^2 (r_2-r_1)			\right]
	\,.
 \label{31XII15.2}
\eena

The action of the Einstein-Maxwell system is given by
\bena
	S&=& - \frac{1}{16 \pi} \int  (R- 2 \Lambda -F^2) \sqrt{g} \,   d^4 x
		\, \nonumber
\\
     &=& \underbrace{- \frac{1}{16 \pi} \int  (R- 2 \Lambda) \sqrt{g} \,   d^4 x}_{:=S_{G}}
		\,
	+	\underbrace{ \frac{1}{16 \pi} \int  F^2 \sqrt{g} \,   d^4 x }_{:=S_{\mbox{\rm \tiny EM}}}
\,.
\eena
Let $S_G$ be the gravitational action, we have
\bena
	S_{G}&=& - \frac{1}{16 \pi} \int  (R- 2 \Lambda) \sqrt{g} \,   d^4 x
	\, \nonumber
\\
	&=& -\frac{\Lambda}{8 \pi}  V
	 \, \notag
\\
 	&=& -\Lambda \frac{\pi}{3 \kappa  \Xi ^2}\left[\left(r_2^3-r_1^3\right)-a^2 (r_2-r_1)			\right]
 	\,.
  \label{30XII15.2}
\eena
A {\sc Mathematica} calculation gives
\bena
	F^2:=  g^{\alpha\beta} g^{\mu\nu} F_{\alpha\mu} F_{\beta\nu} =\frac{(e-p)^2}{(a \cos (\theta
   )+r)^4}+\frac{(e+p)^2}{(r-a
   \cos (\theta ))^4}
\,,
\eena
leading  to
\bena
	S_{\mbox{\rm \scriptsize EM}}&=&\frac{1}{16 \pi} \int  F^2 \sqrt{g} \,   d^4 x
	\, \nonumber
\\
	&=& \frac{1}{16 \pi}  \frac{4 \pi^2}{\kappa} \int_{r_1}^{r_2} \int_0^\pi \, F^2 \sqrt{g} d \theta d r
	 \, \notag
%\\
%	&=& \frac{\pi}{4 \kappa}  \int_{r_1}^{r_2} \int_0^\pi \, F^2 \sqrt{g} d \theta d r
%	 \,  \nonumber
 \\
	&=&  \frac{\pi (p^2 +e^2)}{ \kappa \Xi^2} \left(\frac{r_1}{r_1^2-a^2}-\frac{r_2}{r_2^2-a^2}\right)
	 \,
	 \,.
\eena
Together this yields
\bena
	S=\frac{\pi}{ \kappa \Xi^2} \left\{ -\frac{\Lambda}{3} \left[\left(r_2^3-r_1^3\right)-a^2 (r_2-			r_1)	\right] +   (e^2+p^2 ) \left(\frac{r_1}{r_1^2-a^2}-\frac{r_2}{r_2^2-a^2}\right)  \right\}
	 \,.
\eena
The minimum of the action is attained at $(n_1,n_2)=(2,1)$, and equals  $S_{\min{}}\approx -2.357$.
Since $r_1\to_{n_1\to\infty} a$ and  $\noQ\to_{n_1\to\infty} 0.32$ (see \eq{30XII15.1}), the action is unbounded from above.
It follows from the analysis in Section~\ref{s22VI15.1} that $S_G$ is bounded from above by $-\pi/2$, so only the Maxwell action grows without bound.
Now, if $r_2$ is close to $r_1$, then the Maxwell action is very small. One expects this to be true when both $n_1$ and $n_2$ are very large. This suggests very strongly that the set of pairs  $(n_1,n_2)$, for which the Maxwell action  $S_{\mbox{\rm \scriptsize EM}}$ is very small compared to the gravitational one, is unbounded. Numerics shows that this is indeed the case for all large numbers $n_1$ that we have looked at.

In particular  solutions with very large values of $n_1-n_2$ are strongly suppressed  when path-integral  arguments are invoked.

\subsection{Lorentzian case}

In this section we consider the \emph{Lorentzian solutions with $e=0$ and with the value of $a$, $M$ and $\noQ$ arising from a smooth compact Euclidean solution with $e=0$}.
To avoid ambiguities, we write
\bel{2I16.1}
 \Delta_{\mathrm{Lor}} :=(r^2+a^2)(1- \lambda r^2)-2Mr + p^2+e^2
 \, \ \textrm{
and } \ \Xilor := 1 + \lambda a^2
 \,.
\ee

\begin{figure}
\begin{center}
{\includegraphics[scale=.9,angle=0]{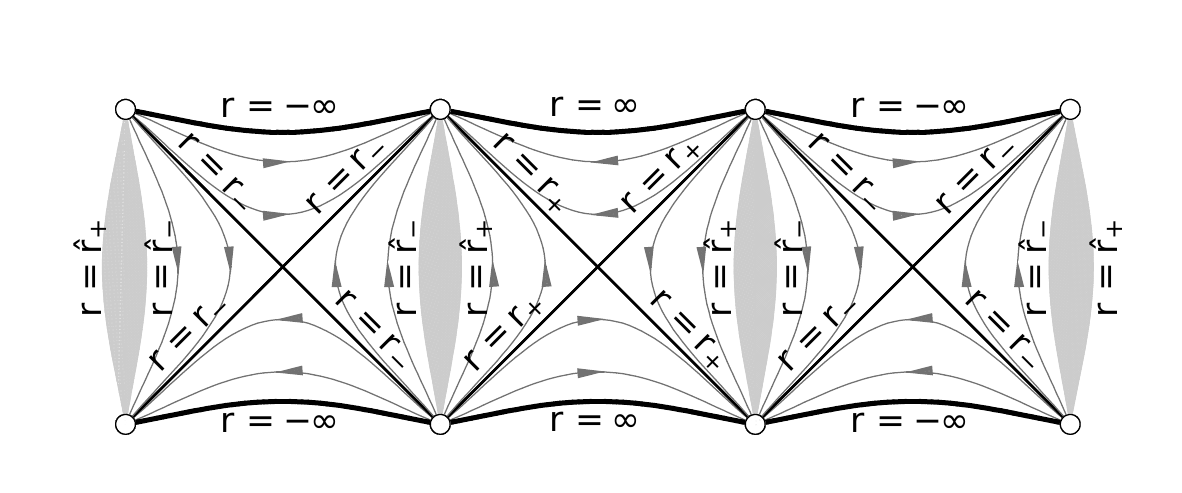}}
\caption{A projection diagram for the Kerr-Newman - de Sitter metrics with exactly two distinct real first-order zeros of $\Delta_r$, $r_-<0<r_+ $, from~\cite{COS}. Outside of the shaded regions, which contain the singular rings and the time-machines with boundaries at $\hat r_\pm$, the diagram represents   accurately (within the limitations of a two-dimensional projection) the global structure of the space-time. Here $r_-$ and $r_+$ indicate the radii of the Lorentzian horizons, not to be confused with the Euclidean rotation axes from the body of the paper.}
\label{F25VI12.7}
\end{center}
\end{figure}

In all solutions that we have found the function $\Deltalor$  has precisely two real first-order zeros, with exactly one positive, denoted by $r_+$. The associated horizon is usually
referred to as the \emph{cosmological horizon}. The global structure of the Lorentzian solution is shown in Figure~\ref{F25VI12.7}.

As already pointed-out, there is an ambiguity in the definition of total mass of the associated Lorentzian space-time. In a Hamiltonian approach this ambiguity is related to the choice of the Killing vector field for which we calculate the Hamiltonian~\cite{CJKKerrdS}.
In any case, the physical mass $M_{\phys}$ and the angular momentum $J_{\phys}$ are usually calculated using the formulae
\bena
	M_{\phys}= \frac{M}{ \Xilor ^2} \,, \quad
 	J_{\phys}= \frac{a M}{ \Xilor ^2}
 %\,, \quad
 %	\ephysphys{}= \frac{\sqrt{\noQ}}{ \Xilor }
  \,.
 %\,, \quad
\label{15VI19.1}
\eena
(The above mass of the Lorentzian solution is obtained by calculating the Hamiltonian associated with the Killing vector field $\Xilor \partial_t +3^{-1} \Lambda a \partial_\varphi$, while the total angular momentum is the Hamiltonian associated with $\partial_\varphi$.)

The area of the cross-section of the horizon located at $r_+$ is given by
\bena
	A_+ = \frac{4 \pi  \left(r_+^2+a^2\right)}{\Xilor  }
    \,,
\eena
and is usually interpreted as the entropy of the cosmological horizon~\cite{GibbonsHawkingCEH}.
The surface gravity of the horizon $r=r_+$ associated with the Killing vector $X:= \partial_t + \Omega \partial_\varphi$, where $\Omega$ is chosen so that $X$ is tangent to the generators of the horizon, is
\bena
	\kappa_+=\frac{1}{2 \Xilor  (r_+^2+a^2)} \Deltalor  ' |_{r={r_+}} \,.
\eena
%

%%%%%%%%%%%%%%%%%%%%%%%%%%%%%%%%%%%%%%%%%%%%%%%%%%%%%%%%%%%%%%%%%%%%%%%%%%%%%%%%%%%%%%%%%%%%%%%

%%%\input{overview_newCH}%%%%%%%%%%%%%%%%%%%%%%%%%%%%%%%%%%%%%%%%%%%%%%%%%%%%%%%%%%%%%%%%%%%%%%
\section{A sample}

We list in Table~\ref{T15IX15.1} the defining parameters of all solutions for $\lambda=1,\;\zeta =-1\,,n_1,n_2 \in \{-10,10\}\,, n_1 > n_2$,
fulfilling the constraints, as well as some associated physical quantities.
The constraints $a<r_1<r_2$ and $a^2<1$ are clearly seen to be fulfilled.
The physical quantities $ M_{\textrm{phys}}$, $|J_{\phys}| $, $\ephysphys $ are defined in \eq{15VI19.1}, while ${S}$ denotes the Euclidean action of the solutions.
\flushleft

\begin{table}[h!]
\bena
\begin{tiny}
\begin{array}{ccccccccccccc}
 n_1 & n_2 & n_1-n_2 & \text{a} & r_1 & r_2 & \text{M} & \esquare{} & M_{\textrm{phys}} &|J_{\phys}| & \ephysphys & S_{G} & S \\
 2 & 1 & 1 & 0.05720 & 0.4147 & 0.5837 & 0.2449 & 0.06344 & 0.2433 & 0.01392 & 0.2511 & -2.368 & -2.357 \\
 3 & 1 & 2 & 0.09837 & 0.3698 & 0.6253 & 0.2398 & 0.06764 & 0.2352 & 0.02314 & 0.2576 & -2.377 & -2.335 \\
 3 & 2 & 1 & 0.05939 & 0.4494 & 0.5488 & 0.2497 & 0.06610 & 0.2480 & 0.01473 & 0.2562 & -2.353 & -2.340 \\
 4 & 1 & 3 & 0.1264 & 0.3426 & 0.6493 & 0.2366 & 0.07260 & 0.2292 & 0.02898 & 0.2652 & -2.380 & -2.291 \\
 4 & 2 & 2 & 0.1063 & 0.4159 & 0.5785 & 0.2504 & 0.07462 & 0.2449 & 0.02604 & 0.2701 & -2.341 & -2.290 \\
 4 & 3 & 1 & 0.05997 & 0.4638 & 0.5344 & 0.2510 & 0.06681 & 0.2492 & 0.01494 & 0.2576 & -2.349 & -2.336 \\
 5 & 1 & 4 & 0.1460 & 0.3247 & 0.6646 & 0.2346 & 0.07709 & 0.2249 & 0.03284 & 0.2718 & -2.380 & -2.233 \\
 5 & 2 & 3 & 0.1403 & 0.3930 & 0.5971 & 0.2518 & 0.08399 & 0.2422 & 0.03398 & 0.2842 & -2.326 & -2.211 \\
 5 & 3 & 2 & 0.1088 & 0.4370 & 0.5571 & 0.2538 & 0.07688 & 0.2479 & 0.02698 & 0.2740 & -2.330 & -2.276 \\
 5 & 4 & 1 & 0.06020 & 0.4717 & 0.5265 & 0.2515 & 0.06710 & 0.2497 & 0.01503 & 0.2581 & -2.347 & -2.334 \\
 6 & 1 & 5 & 0.1602 & 0.3120 & 0.6751 & 0.2333 & 0.08086 & 0.2217 & 0.03553 & 0.2772 & -2.380 & -2.165 \\
 6 & 2 & 4 & 0.1648 & 0.3768 & 0.6096 & 0.2533 & 0.09238 & 0.2401 & 0.03957 & 0.2959 & -2.312 & -2.110 \\
 6 & 3 & 3 & 0.1452 & 0.4173 & 0.5721 & 0.2571 & 0.08814 & 0.2466 & 0.03580 & 0.2908 & -2.308 & -2.182 \\
 6 & 4 & 2 & 0.1100 & 0.4492 & 0.5447 & 0.2552 & 0.07789 & 0.2492 & 0.02740 & 0.2758 & -2.325 & -2.270 \\
 6 & 5 & 1 & 0.06032 & 0.4767 & 0.5215 & 0.2518 & 0.06724 & 0.2499 & 0.01508 & 0.2584 & -2.346 & -2.333 \\
\end{array}
\end{tiny}
\nn
\eena
\caption{\label{T15IX15.1} Some selected solutions with the most relevant physical parameters
in dimensionless units}
\end{table}
%

%%%%%%%%%%%%%%%%%%%%%%%%%%%%%%%%%%%%%%%%%%%%%%%%%%%%%%%%%%%%%%%%%%%%%%%%%%%%%%%%%%%%%%%%%%%%%%%

%%%\input{si_newCH}%%%%%%%%%%%%%%%%%%%%%%%%%%%%%%%%%%%%%%%%%%%%%%%%%%%%%%%%%%%%%%%%%%%%%%%%%%%%
\section{SI units}
 \label{SI}

Recall that $\lambda:= \Lambda / 3$. The replacements
\be
 r \mapsto \sqrt{\frac{1}{\lambda}}\times  r
  \,,
  \quad
 M \mapsto \sqrt{\frac{1}{\lambda}}\times  M
  \,,
  \quad
 a \mapsto \sqrt{\frac{1}{\lambda}}\times  a
  \,,
    \quad
 e \mapsto \sqrt{\frac{1}{\lambda}}\times e
   \,,
\ee
yield
\bena
    \Delta_r \mapsto \frac{1}{\lambda} \Big(\underbrace{ (r^2+a^2)\left( 1-  r^2 \right) -2 M r +\esquare{}}_{:= \Delta^{\lambda=1}_r} \Big)
     \,,
    \quad
    \Delta'_r \mapsto \frac{1}{\sqrt{\lambda}}  \Delta'^{\lambda=1}_r
      \,,
    \quad
     \Xi \mapsto \underbrace{1-a^2}_{:=\Xi ^{\lambda=1}}
      \,.
\notag \\
\eena
It is easy to check that if
\bena
\Big(r_1^{\lambda=1},r_2^{\lambda=1},M^{\lambda=1},a^{\lambda=1},(\esquare{})^{\lambda=1}  \Big)
\notag
\eena
is a solution of the system Eq.(\ref{15I12.0}-\ref{15I12.4}) for $\lambda=1$, then
\bena
  r_1 &=& \sqrt{\frac{1}{\lambda}} \times  r_1^{\lambda=1}
     \,,
 \quad
  r_2 = \sqrt{\frac{1}{\lambda}} \times  r_2^{\lambda=1}
     \,,
 \quad
  M = \sqrt{\frac{1}{\lambda}} \times M^{\lambda=1}
     \,,
\nn \\
  a &=& \sqrt{\frac{1}{\lambda}} \times  a^{\lambda=1}
     \,,
 \quad
  \esquare{} = \frac{1}{\lambda} \times  (\esquare{})^{\lambda=1}
     \,,
\eena
provides a solution of this system with an arbitrary value $\lambda$.

In SI-units we have
\be
 M_{\phys}^{SI}= \frac{c^2}{G}  \times  M_{\phys}
  \,,
  \quad
 \ephysphys  ^{SI}= \sqrt{\frac{4 \pi \epsilon_0 c^4}{G} } \times   \ephysphys
  \,,
\ee
where $G$ is the gravitational constant, $c$ the speed of light and $\epsilon_0$  the
electric constant.
Then the physical angular momentum in SI-units can be computed as
\bena
  J_{\phys}^{SI} &=& a \times c \times M_{\phys}^{SI}  \;.
\eena
Putting all this together we obtain
\bena
   M_{\phys}^{SI} &=&  \frac{c^2}{G}  \times  \frac{1}{\Xilor ^2} \times  \sqrt{\frac{1}                                                                                                 {\lambda}} \times  M^{\lambda=1}
     \,,
 \\
  a^{SI} &=& \sqrt{\frac{1}{\lambda}} \times a^{\lambda=1}
     \,,
 \\
   \ephysphys{}^{SI} &=& \sqrt{\frac{4 \pi \epsilon_0 c^4}{G} \times \frac{1}{\lambda} \times  \frac{1}{\Xilor ^2}  \times (\esquare{})^{\lambda=1}}
     \,,
 \\
  J_{\phys}^{SI} &=& c \times a^{SI} \times  M_{\phys}^{SI} \;.
\eena
Since $\Xi_{\mathrm{{Lor}}}$ and $\Delta_{\mathrm{{Lor}}}$ are invariant under rescaling, it follows
\bena
	A_+^{SI}&=& \frac{1}{\lambda} A_+^{\lambda=1}
     \,,
 \\
	\kappa_+^{SI}&=& c^2 \sqrt{\lambda} \kappa_+^{\lambda=1}
	\,.
\eena

The black hole temperature in SI-units $T^{SI}$ reads
\bena
	T_{kg^{-1}}&=& \frac{1}{2 \pi} \frac{G}{c^2} \sqrt{\lambda} \times \kappa_+^{\lambda=1}
     \,,
 \\
	T^{SI}&=& \frac{c^3 \hbar}{k \, G} \times T_{kg^{-1}}
	\,,
\eena
where $\hbar= 1.054 \times 10^{-34} \, Js$ and $k=1.38 \times 10^{-23} J K^{-1}$ are the reduced Planck's constant and the Boltzmann constant respectively.
Table \ref{T9IX15.2} lists some values of $M_\phys$ in units of the mass of Milky Way, taken to be $ 10^{12} M_\odot$,
 $$
  M_\phys = \Mastro  \times 10^{12} M_\odot \sqrt{\frac{\Lambda}{\Lambda_0}}
   \,,
 $$
 where $M_\odot$ is the mass of the sun and
 $$
  \Lambda_0 = 3 H_0^2 \Omega_\Lambda = 1.11  \times 10^{-52} \; m^{-2}
 $$
  is the value of the cosmological constant as resulting from the Planck observations~\cite{Planck16.2013} (compare~\cite{Planck13,Supernova2004,Komatsu:2010fb}). We moreover use
 $
  G =  6.67  \times 10^{-11} \; \frac{m^3}{kg s^2}$,
    $
 c = 299 \times 10^{8} \; \frac{m}{s}
    $, and $
   \epsilon_0 =  8.85  \times \; 10^{-12} m^{-3} kg^{-1} s^4 A^2
    $,
$M_{_\odot} = 1.99 \times 10^{30}\;kg$.
\renewcommand{\arraystretch}{1.2}
\renewcommand{\arraystretch}{1.2}
\begin{table}[h!]
\begin{tabular}{ | c | c | }
\hline  $ M_{\phys} / 10^{10} M_{\textrm{Milky Way}} $ & \text{type}
\\
  \hline
 $2.27   $ & \text{minimum}
\\
  $2.84  $  & \text{ maximum}
\\
  $2.71  $  & \text{at minimal charge}
\\ $2.77  $ & \text{at maximal charge} \\ \hline
\end{tabular}
\caption{\label{T9IX15.2}
 Some cosmological values of $M_\phys$, in Milky Way mass units.}
\end{table}

Another set of amusing questions is, which values of $\Lambda$ are required
to obtain the charge $e^-$ of an  electron  as minimal value for the physical charge $ \ephysphys{}^{SI}=e^-$, or
the mass of an electron $M_{\mathrm{e}}$, or of a proton $M_{\mathrm{p}}$,  as minimal value of the physical mass:
\bena
e^-= 1.60 \times 10^{-19} C
\,, \quad
M_{\mathrm{e}}= 9.11 \times 10^{-31} kg
\,, \quad
M_{\mathrm{p}}= 1.67 \times 10^{-27} kg
\,.
\nn
\eena
The results are given in Table \ref{T9IX15.12}.
\begin{table}[h!]
\begin{tabular}{ | c | c | c |}
\hline  \text{minimal physical mass /charge} & $\Lambda$ / $m^{-2}$ & $\Lambda$ / $\Lambda_0$
\\
  \hline
 $e^-$ & $9.92 \times 10^{70}$ & $8.94 \times 10^{122} $
\\
  $M_{\mathrm{e}}$  & $2.72 \times 10^{113}$ & $2.45 \times 10^{165} $
\\
  $M_{\mathrm{p}}$ & $8.06 \times 10^{106}$
   & $7.26 \times 10^{158}$
    \\ \hline
\end{tabular}
\caption{\label{T9IX15.12}
Values of $\Lambda$ required to obtain $e^-$ as minimal physical charge and
$M_{\mathrm{e}}$/ $M_{\mathrm{p}}$ as minimal physical mass. $\Lambda_0$ is the current estimate of the value of the cosmological constant.}
\end{table}

%%%%%%%%%%%%%%%%%%%%%%%%%%%%%%%%%%%%%%%%%%%%%%%%%%%%%%%%%%%%%%%%%%%%%%%%%%%%%%%%%%%%%%%%%%%%%%%

%%%\input{lor_newCH}%%%%%%%%%%%%%%%%%%%%%%%%%%%%%%%%%%%%%%%%%%%%%%%%%%%%%%%%%%%%%%%%%%%%%%%%%%%
\section{Lorentzian partner solutions}

Consider a set of parameters $n_1$, $n_2$, $M$, $a$, and $\esquare{}$ that
solve, together with the positive zeros of $\Delta_r$, the system
(\ref{15I12.0}-\ref{15I12.4}) and fulfill the constraints. For this
set of parameters  we calculate the zeros of the Lorentzian partner $\Deltalor$ given by \eq{2I16.1}
of the Euclidean function $\Delta_r$. As already mentioned, for all $(n_1,n_2)$ that we have investigated the function  $\Deltalor $ has only two real first-order zeros, with exactly one   positive zero $r_+$.

\subsection{Geometric units}
 \label{ss31XII15.1}

In Table~\ref{T14IX15.1} we list the values of $r_+$, the surface gravity (``temperature'') and the area (``entropy'') of the horizon.
\begin{table}[h!]
\bena
\begin{tiny}
\begin{array}{cccccc}
 n_1 & n_2 & n_1-n_2 & r_+ & \kappa _+ & A_+ \\
 2 & 1 & 1 & 0.612 & -0.246 & 4.730 \\
 3 & 1 & 2 & 0.667 & -0.377 & 5.663 \\
 3 & 2 & 1 & 0.594 & -0.216 & 4.467 \\
 4 & 1 & 3 & 0.699 & -0.452 & 6.239 \\
 4 & 2 & 2 & 0.649 & -0.355 & 5.368 \\
 4 & 3 & 1 & 0.589 & -0.208 & 4.394 \\
 5 & 1 & 4 & 0.719 & -0.498 & 6.617 \\
 5 & 2 & 3 & 0.682 & -0.439 & 5.967 \\
 5 & 3 & 2 & 0.643 & -0.349 & 5.282 \\
 5 & 4 & 1 & 0.587 & -0.204 & 4.364 \\
 6 & 1 & 5 & 0.732 & -0.529 & 6.880 \\
 6 & 2 & 4 & 0.703 & -0.493 & 6.376 \\
 6 & 3 & 3 & 0.676 & -0.438 & 5.890 \\
 6 & 4 & 2 & 0.641 & -0.346 & 5.244 \\
 6 & 5 & 1 & 0.586 & -0.203 & 4.349 \\
\end{array}
\end{tiny}
\nn
\eena
\caption{\label{T14IX15.1} The surface gravity and area for some selected solutions, with $\Lambda=3$.}
\end{table}

\subsection{SI units,  $\Lambda= 1.11 \times 10^{-52} \; m^{-2}$}

With the formulae given in Appendix~\ref{SI} we can calculate the interesting physical
quantities in SI-units for the measured cosmological value $\Lambda_0$ of $\Lambda$  from the data for
$\Lambda=3$. Using the
Planck mission data $\Omega_{\Lambda}=0.6911$ and $H_0=67.74 \, km/(s \, M pc)$,   (see~\cite{Planck13}, p.~31, TT, TE, EE + lowP + lensing), the cosmological constant can be calculated to be
\bena
 \Lambda_0 c^2 =3 \,  H_0^2 \, \Omega_{\Lambda} =  9.99\times  10^{-36} s^{-2} \Rightarrow \Lambda_0= 1.11 \times 10^{-52} \; m^{-2}
	\; \notag
\eena
The reader will find some physical quantities of interest associated with our solutions in Tables~\ref{T14IX15.3} and \ref{T14IX15.2}.
\begin{table}[th]
\bena
\begin{tiny}
\begin{array}{cccccccccc}
 n_1 & n_2 & n_1-n_2 & r_{+} / 10^{26} m & M_{phys} / 10^{52} kg & |J_{phys}| / 10^{86} kg m^2 s^{-1}   & |q_{phys}|  / 10^{42} C & |\kappa_+| / 10^{-10} m s^{-2} & A_+  / 10^{53} m^{2}
 & T  / 10^{-30} K  \\
 2 & 1 & 1 & 1.006 & 5.387 & 1.519 & 4.790 & 1.345 & 1.278 & 0.545 \\
 3 & 1 & 2 & 1.097 & 5.207 & 2.524 & 4.914 & 2.062 & 1.530 & 0.836 \\
 3 & 2 & 1 & 0.977 & 5.490 & 1.607 & 4.888 & 1.181 & 1.207 & 0.479 \\
 4 & 1 & 3 & 1.149 & 5.074 & 3.162 & 5.060 & 2.470 & 1.686 & 1.001 \\
 4 & 2 & 2 & 1.066 & 5.421 & 2.841 & 5.153 & 1.938 & 1.451 & 0.786 \\
 4 & 3 & 1 & 0.969 & 5.517 & 1.631 & 4.914 & 1.136 & 1.188 & 0.461 \\
 5 & 1 & 4 & 1.181 & 4.978 & 3.583 & 5.186 & 2.723 & 1.788 & 1.104 \\
 5 & 2 & 3 & 1.120 & 5.362 & 3.708 & 5.422 & 2.401 & 1.613 & 0.974 \\
 5 & 3 & 2 & 1.057 & 5.487 & 2.944 & 5.228 & 1.906 & 1.427 & 0.773 \\
 5 & 4 & 1 & 0.966 & 5.528 & 1.640 & 4.924 & 1.117 & 1.179 & 0.453 \\
 6 & 1 & 5 & 1.203 & 4.909 & 3.877 & 5.289 & 2.891 & 1.859 & 1.172 \\
 6 & 2 & 4 & 1.156 & 5.316 & 4.318 & 5.645 & 2.696 & 1.723 & 1.093 \\
 6 & 3 & 3 & 1.112 & 5.459 & 3.906 & 5.547 & 2.393 & 1.592 & 0.970 \\
 6 & 4 & 2 & 1.053 & 5.517 & 2.990 & 5.261 & 1.893 & 1.417 & 0.768 \\
 6 & 5 & 1 & 0.964 & 5.534 & 1.645 & 4.929 & 1.108 & 1.175 & 0.449 \\
\end{array}
\end{tiny}
\nn
\eena
\caption{\label{T14IX15.3} Some physical quantities in SI units for
selected solutions}
\end{table}
\begin{table}[th]
\bena
\begin{tiny}
\begin{array}{cccccccccccccccc}
 n_1 & 2 & 3 & 3 & 4 & 4 & 4 & 5 & 5 & 5 & 5 & 6 & 6 & 6 & 6 & 6 \\
 n_2 & 1 & 1 & 2 & 1 & 2 & 3 & 1 & 2 & 3 & 4 & 1 & 2 & 3 & 4 & 5 \\
 \frac{M_{phys}}{M_{\odot}} / 10^{22} & 2.708 & 2.618 & 2.760 & 2.551 & 2.726 & 2.774 & 2.503 & 2.696 & 2.759 & 2.779 & 2.468 & 2.672 & 2.745 & 2.774 & 2.782 \\
=\frac{M_{phys}}{M_{gal}}  / 10^{10} &  &  &  &  &  &  & &  &  &  &  &  &  &  &  \\
\end{array}
\end{tiny}
\nn
\eena
\caption{\label{T14IX15.2} The physical mass  in solar mass- and galaxy mass units for some
selected solutions}
\end{table}

To close this section, let us assume that the above  universe  consists of protons,
neutrons, and hydrogen atoms.
 This means that for the range of values, as given above, we have $n_{\mbox{\scriptsize \rm items}}\approx M_{\phys}/M_{\mbox{\scriptsize \rm proton}} \approx 2 \times 10^{79}$ items. On the other hand $n_{\mbox{\scriptsize \rm protons}}=|q_{\phys}|/e^- \approx 2 \times 10^{61}$ particles are required to produce the required charge. As a consequence, every $10^{18}$-th item carries a charge.

%%%%%%%%%%%%%%%%%%%%%%%%%%%%%%%%%%%%%%%%%%%%%%%%%%%%%%%%%%%%%%%%%%%%

%%%\input{pl_new2CH_mod}%%%%%%%%%%%%%%%%%%%%%%%%%%%%%%%%%%%%%%%%%%%%%%%%%%%%%%%%%%%%%%%%%%%%%%%
\flushleft
\section{Page limit}
 \label{s7IX15.1}

The aim of this appendix is to discuss the charged solutions obtained by Page's limiting procedure~\cite{PageInstanton}.  (These solutions have been already been discussed in~\cite[Section~7.4,  Equations~(135)-(136)]{KunduriLuciettiLR} from a rather different perspective; compare~\cite{MellorMoss}.)
Recall that Page's approach is the following: Let $r_0$ be a zero of $\Delta_r$, and let $\epsilon$ be a small parameter.
We define new  coordinates $(\chi,\barvarphi , \eta)$ as
\bena
	r&=&r_0-\epsilon  \cos (\chi )\label{10I16.1}
\,,
\\
    \varphi &=& \barvarphi  -\frac{a}{r_0^2-a^2}t \label{10I16.2}
\,,
\\
     t &=& \frac{\omega_0\eta}{\epsilon} \label{10I16.3}
\,,
\eena
where  $\eta$ and
$\barvarphi $ are $2\pi$-periodic,  and  $\omega_0$ is a constant to be determined.
We choose the parameters $(M,a,\esquare)$ so that
\bena
	\Delta_r &= & C \left(1-\cos ^2(\chi )\right) \epsilon ^2 + O( \epsilon^3 )  \,,
\eena
for a suitable constant $C=C(\epsilon)$.
After taking the limit $\epsilon \to 0$ the metric takes the form
\bena
	ds^2&=&
 3 \left( r_0^2 -a^2 \cos ^2(\theta ) \right) \times
\\
 &&
 \nn
  \Bigg\{
  \frac{1}{ 6 \Lambda r_0^2 -a^2 \Lambda -3} \left( d \chi^2
  + \frac{ \left(6 \Lambda r_0^2 -a^2 \Lambda -3 \right)^2 \omega_0^2}{\left( r_0^2 -a^2 \right)^2 \left(3 -a^2 \Lambda \right)^2 	}  \sin ^2(\chi ) d \eta^2 \right)
\notag
\\
	&\phantom{=}&+ \frac{1}{3 - a^2 \Lambda  \cos ^2(\theta)}
 	 \Bigg [  d \theta^2 +\frac{   \left( r_0^2-a^2 \right)^2 (3 - a^2 \Lambda  \cos 					^2(\theta))^2
	}{ \left(3 -a^2 \Lambda \right)^2  \left(r_0^2 - a^2 \cos ^2(\theta ) \right)^2} \sin ^2(\theta )
    \times
	\notag
\\
	&\phantom{=}&\phantom{+ \frac{1}{3 - a^2 \Lambda  \cos ^2(\theta)}
 	 \Bigg [ }\Big( \dnotdelta \barvarphi  + \frac{2 a  r_0 \omega_0}{  \left(r_0^2-a^2\right)^2 } \cos (\chi ) \dnotdelta\eta \Big)^2 \Bigg ] \Bigg \}
\,.
\eena
An Euclidean signature will be obtained if
\bel{5VIII15.15}
 a^2 < r_0^2
  \,,
  \quad
 \Lambda a^2 < 3
  \,,
  \quad
    6 \Lambda  r_0^2 -a^2 \Lambda-3 >0
     \,.
\ee
Note that the transformation  $\eta \mapsto -\eta$ has the effect of changing the sign of $a r_0$, so without loss of generality
we can assume that $ar_0>0$. Since a simultaneous change of sign of $a$ and $r_0$ leaves the metric invariant, we can assume
that
$$
 a\ge 0 \ \mbox{ and } \ r_0 >0
 \,.
$$

Near $ \chi=0$ we introduce a new coordinate $\phinotone $, $2\pi$-periodic, chosen so that $g_{\eta\eta}|_{ \chi =0}=0$:
\bel{5VIII15.11}
 \dnotdelta \phinotone  := \alphanotone  \bigg(\dnotdelta \barvarphi  + \frac{2 a  r_0 \omega_0}{  \left(r_0^2-a^2\right)^2 }   \dnotdelta\eta
  \bigg)
 \,,
\ee
for some constant $\alphanotone \in\R^*$. Standard considerations show that the metric will be smooth if
\bel{5VIII15.12}
 	\displaystyle
 \omega_0^2 \frac{ \left(6 \Lambda r_0^2 -a^2 \Lambda -3 \right)^2 }{\left( r_0^2 -a^2 				\right)^2 \left( 3 -a^2 \Lambda  \right)^2 	}=1
 \quad
 \Longleftrightarrow
 \quad
 \omega_0 = \pm  \underbrace{\frac{\left( r_0^2 -a^2 				\right)  \left(3-a^2 \Lambda \right)  	}{ 6 \Lambda r_0^2 -a^2 \Lambda -3  }
  }_{=:\omega_P>0}
 \,,
\ee
\be
   \displaystyle \alphanotone ^2
    \frac{   \left( r_0^2-a^2 \right)^2 ( a^2 \Lambda  \cos 										^2(\theta)-3)^2
	}{ \left( 3 -a^2 \Lambda \right)^2  \left(r_0^2 - a^2 \cos ^2(\theta ) \right)^2}\bigg|_{\Theta = 0} = 1
 \label{5VIII15.13}
 \quad
 \Longleftrightarrow
 \quad
	\alphanotone   = \pm 1
\,.
\ee
(The constant $\omega_P$ of \eq{5VIII15.12}  coincides with Page's constant $\omega_{\mathrm{Page}}$,
\bena
\omega_{\mathrm{Page}} =    \frac{ r_0^2 (3-a^2 \Lambda  )  \left(r_0^2-a^2\right)}{3(a^2 +\Lambda  r_0^4)}
 \,,
\eena
when $\noQ=0$ and
when the requirement that $r_0$ is a double zero of $\Delta$, which is implicit in the construction here, is taken into account.)

When $a=0$, the metric is now a product of two round metrics, with possibly different curvatures, on $S^2 \times S^2$. From now on
we only consider the case
$$
 a>0
 \,.
$$

Near $ \chi=\pi $ we introduce a new angular coordinate $\hat \phinotone $, $2\pi$-periodic, chosen so that $g_{\eta\eta}|_{ \chi =\pi}=0$:
\bel{22I16.5}
 \dnotdelta \hat \phinotone  := \hat \alphanotone
  \bigg(\dnotdelta \barvarphi  - \frac{2 a  r_0 \omega_0}{  \left(r_0^2-a^2\right)^2 }  \dnotdelta\eta
  \bigg)
 \,.
\ee
One checks that smoothness of the metric there is already guaranteed by \eq{5VIII15.12}-\eq{5VIII15.13}.

Eliminating $\dnotdelta \barvarphi$ between \eq{5VIII15.11} and \eq{22I16.5} we find
\bel{25I16.1}
 \hat \alphanotone \dnotdelta \hat \phinotone   = \alphanotone  \dnotdelta \phinotone  - \frac{4 a  r_0 \omega_0}{  \left(r_0^2-a^2\right)^2 }   \dnotdelta\eta
 \,,
\ee
Keeping in mind that $\eta$, $\phinotone $ and $\hat \phinotone $ are $2\pi$-periodic, we are led to the condition
\bel{5VIII15.1}
  \frac{4  a  r_0  \omega_0 }{  \left(r_0^2-a^2\right)^2 } =:   n \in \Z^*
 \,.
\ee
Equivalently,
\bel{28X15.1}
\frac{4  a  r_0 \left( 3 - a^2 \Lambda  \right)}{  \left(r_0^2-a^2\right) \left(6 \Lambda r_0^2 -a^2 \Lambda -3 \right)} = |n|  \in \N^*
\,.
\ee
To proceed, we prescribe $  n \in \Z^*$, 
solve the system  consisting of the equations $\Delta(r_0)=\Delta'(r_0)$ together with \eq{28X15.1}
for $(r_0,a,M)$, and check if the constraints are fulfilled.

\subsection{Parametrization of $r_0$ and $a$ by $\nu$ and $\bar e$}

One can provide an explicit parameterisation of  solutions of the equations
\beal{11VIII15.1}
 	\Delta_r(r_0,a,M,\esquare{})=0
\,,
\\						
 	\Delta'_r(r_0,a,M)=0
\,,
\eeal{11VIII15.2}
which proceeds as follows: Solving (\ref{11VIII15.2}) for $M$ yields
\bel{11VIII15.3}
	M=\frac{1}{3} r_0 \left(a^2 \Lambda -2 \Lambda  r_0^2+3\right)
\,.
\ee
Using (\ref{11VIII15.3}) in (\ref{11VIII15.1})  and introducing $\nu\in (0,1)$
and $\bar e \in \R$   through the equations
$$
 \mbox{$a=\nu r_0$ and $\esquare{}= \bar e r_0^2$}
$$
(note that $r_0\ne 0$ by \eq{5VIII15.15}, and  that we allow now a negative $\esquare = p^2-e^2$)
 leads to
\bel{11VIII15.5}
	\left(1-\nu^2\right) \left(1-\frac{\Lambda  r_0^2}{3}\right)-\frac{2}{3}  \left(\nu^2 r_0^2 			\Lambda -2 \Lambda  r_0^2+3\right)+ \bar  e=0
\,.
\ee
Solving (\ref{11VIII15.5}) for $r_0$, one is led to the condition
\bel{29X15.1}
    \bar e< 1+\nu^2
    \,,
\ee
together with
\bel{11VIII15.6}
	r_0=   \sqrt{\frac{3 (\nu ^2+1  -   \bar e)}{3-\nu ^2}} \frac{1}{\sqrt{\Lambda }}
\,.
\ee
(\ref{11VIII15.6}) and $a= \nu r_0$ inserted in (\ref{11VIII15.3}) give
\bel{6VIII15.25}
	M=\frac{\left(1-\nu ^2\right)^2+ \bar e (2-\nu^2)}{3-\nu ^2} r_0
\,.
\ee
Using (\ref{11VIII15.6}) and $a=\nu r_0$ in (\ref{28X15.1})  yields
\bel{30X15.1}
	n= \frac{4\nu  \left(( \bar e -2) \nu ^2-\nu ^4+3\right)}{\left(1-\nu ^2\right) \left(( \bar e 		+6) \nu ^2-6 \bar e -\nu ^4+3\right)}
 \,.
\ee
This equation is invariant under the replacement $(n,\nu) \to (-n,-\nu)$ . Hence, from now on we assume
$$
 n>0
 \,.
$$
The constraints  (\ref{5VIII15.15}) then become 
\bel{30X15.2}
 0 < \nu  < 1
  \,,
  \quad
  0 <\nu ^6 - ( \bar e +1) \nu ^4 + 3 (\bar e -3) \nu ^2+9
  \,,
  \quad
  0 <    ( \bar e +6) \nu ^2-6 \bar e -\nu ^4+3
     \,.
\ee

\subsubsection{Magnetic charge equal to electric charge (possibly zero)}

When  $\bar e=0$  the metric coincides with the Page metric, let us  discuss this case for completeness. Equations~(\ref{30X15.1}) and (\ref{30X15.2}) reduce to
\bel{30X15.3}
	n=\frac{4 \nu  \left(\nu ^2+3\right)}{3 +6 \nu ^2-\nu ^4}
\,,
\ee
and
\bel{30X15.4}
 0<\nu < 1
  \,,
  \quad
  0 < \nu ^6-\nu ^4-9 \nu ^2+9
  \,,
  \quad
   0 <    6 \nu ^2-\nu ^4+3
     \,.
\ee

If follows easily, that if the first inequality in (\ref{30X15.4}) holds,
the other two inequalities hold as well. A simple analysis of (\ref{30X15.3}) shows,
that $0<\nu<1$ and $n \in \N^*$ imply $n=1$. For this value of $n$, \eq{30X15.3} can be solved exactly. The only solution fulfilling $0<\nu<1$ is
\bena
	\nu_{Page}&=&-\sqrt{\sqrt[3]{1+\sqrt{2}}-\frac{1}{\sqrt[3]{1+\sqrt{2}}}+2}
 \nn
\\
  &&
  +\frac{1}{2} 			\sqrt{-4 \sqrt[3]{1+\sqrt{2}}+\frac{4}{\sqrt[3]{1+\sqrt{2}}}+\frac{32}{\sqrt{\sqrt[3]{1+		\sqrt{2}}-\frac{1}{\sqrt[3]{1+\sqrt{2}}}+2}}+16}-1
\notag
\\
	& \approx & 0.2817
\,. \label{6VIII15.28}
\eena
Using this value in (\ref{11VIII15.6}) and (\ref{6VIII15.25}) yields
\bena
	r_0=\frac{1.0529}{\sqrt{\Lambda }}
\,,
\quad
	a \approx \frac{0.2967}{\sqrt{\Lambda }}
\,,
\quad
	M \approx \frac{0.3056}{\sqrt{\Lambda }}
\,.
\eena

We continue with the case $\bar e >0$.

\subsubsection{$\bar e >0$}

The addition of a positive charge parameter $\bar e$ increases the right-hand side
of the second inequality in (\ref{30X15.2}) $\forall \nu \in (0,1)$. Thus from
the analysis of the uncharged case, we can conclude that this
constraint holds as well in the charged case.

The right-hand side of the third inequality
in  (\ref{30X15.2}) is monotonously increasing for $  \nu \in (0,1)$. It follows that the infimum and supremum are attained at $\nu=0$ and $\nu=1$ respectively. From this we can conclude the
following:

\begin{itemize}

\item
The inequality $\bar e < \frac{8}{5}$ is a necessary criterion to obtain an Euclidean signature, otherwise  the third constraint in (\ref{30X15.2}) is nowhere satisfied
for $\nu \in (0,1)$.

\item For  $0< \bar e \leq \frac{1}{2}$ (\ref{30X15.2}) is fulfilled  $\forall \nu \in (0,1)$.
A simple analysis of (\ref{30X15.1}), considering the third constraint of (\ref{30X15.2}), shows that $n$ is non-negative and attains every value in $\N$ when $\nu$ varies in $(0,1)$.
Thus if $0< \bar e \leq 1/2$, then for all positive integers $n$ there exists $\nu \in (0,1)$ so that
(\ref{30X15.1}) and (\ref{30X15.2}) are fulfilled.

\item For $\frac{1}{2}< \bar e < \frac{8}{5}$ the right-hand side of the third inequality in (\ref{30X15.2}) has a simple zero at some value $\nu^* \in (0,1)$, thus the constraints (\ref{30X15.2}) are not fulfilled on $(0,\nu^*)$. Futhermore (\ref{29X15.1}) is required.
As the third inequality in (\ref{30X15.2}) is a quadratic in the variable
$\nu^2$, it is easy to verify that this condition holds on $(\nu^*,1)$.
For the interval $(\nu^*,1)$ it follows from a simple analysis that the function which at fixed $\bar e$ assigns to $\nu$ the right-hand side of (\ref{30X15.1}) attains every value in $\N$  above some threshold $n_{\min{}}(\bar e)$ and that the constraints are fulfilled. The zeros of the first derivative of (\ref{30X15.1}) lead to a fifth order polynomial.
Thus the minimum value can only be determined numerically. The result is illustrated in Figure~\ref{F6XI15.1}.
\begin{figure}[h!]
  \begin{center}
    \includegraphics[width=.4 \textwidth]{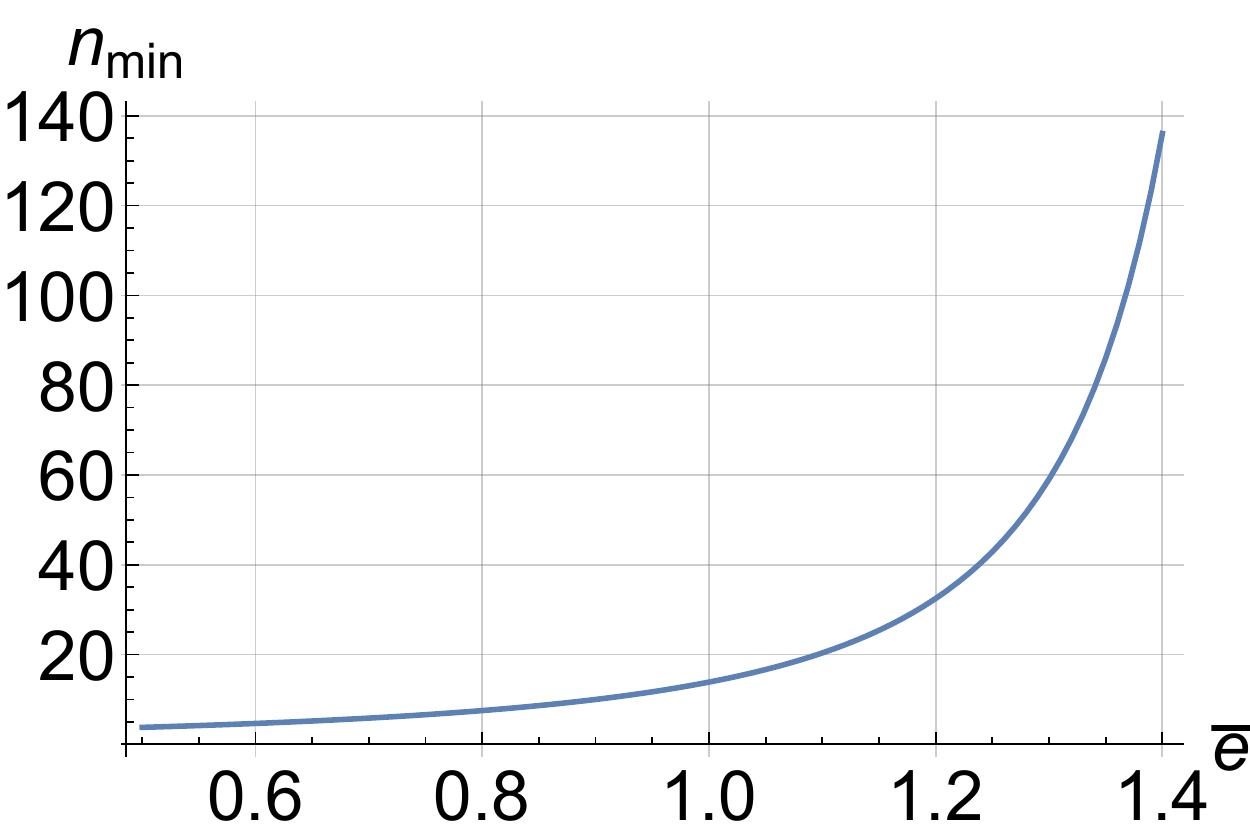}
  \end{center}
  \caption{\label{F6XI15.1} The function $n_{\min{}}(\bar e)$. }
\end{figure}
From the numerical analysis it follows  that $n=4$ is the lowest occurring ``quantum number''
for $\bar e \in \left(\frac{1}{2},\frac{8}{5} \right)$.

\end{itemize}

%\newpage
\subsubsection{$\bar e <0$}

The addition of a negative charge increases the right-hand side
of the third inequality in (\ref{30X15.2}) $\forall \nu \in (0,1)$. Thus from
the analysis of the uncharged case, we can conclude that this
constraint holds as well in the charged case.
The right-hand side of the second inequality
in  (\ref{30X15.2}) is monotonously decreasing in the uncharged case for $  \nu \in (0,1)$ and
attains a zero at $\nu=1$. The addition of a negative charge increases the rate of decreasing. From this it follows that there exists a zero of (\ref{28X15.1}) located at $\nu^* \in (0,1)$. Thus the constraints are fulfilled, for a given negative charge parameter, if and only if  $\nu \in (0,\nu^*)$.

\medskip

\begin{figure}[h!]
  \begin{center}
    \includegraphics[width=.45\textwidth]{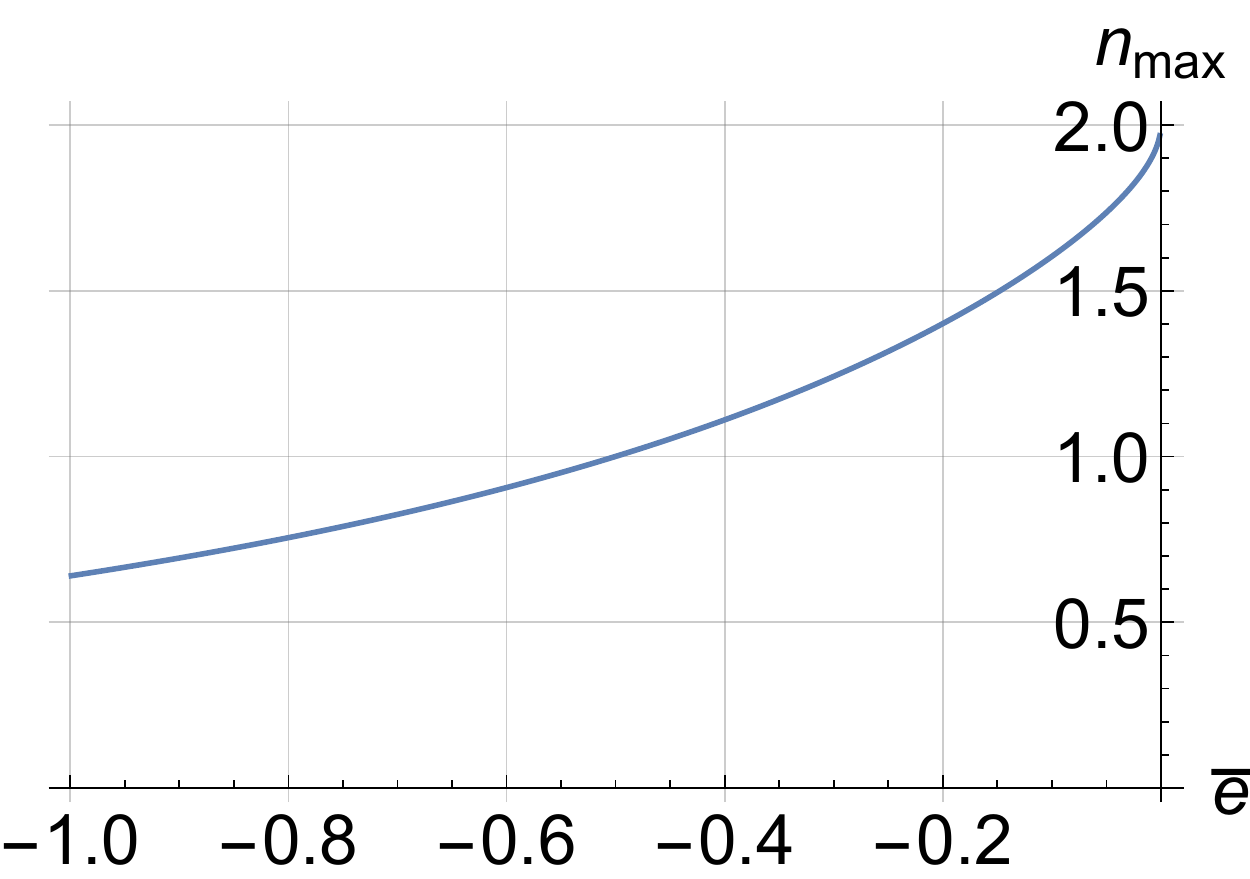}%
  \end{center}
  \caption{\label{F6XI15.2} The function $n_{\max{}}(\bar e)$.}
\end{figure}

The numerator
 of the $n$-function  (\ref{28X15.1}) has no zeros on $(0,\nu^*)$, which follows from the second constraint in (\ref{5VIII15.15}). 
	Thus it suffices to determine if, for a given  parameter $\bar e$, the maximum $n_{\max{}}(\bar e)$ of the function of $\nu$ defined by the right-hand side of  (\ref{30X15.1}), for $\nu\in (0,\nu^*)$, is greater than or equal to one. This analysis can be carried out numerically. The result is illustrated in the plot \ref{F6XI15.2}.
From the numerical analysis we conclude, that $\bar e  \gtrapprox-0.5$
is a necessary
criterion for the existence of a solution, and that $n=1$ is the only possibility when $\bar e \le 0$.
%
% 

%%%%%%%%%%%%%%%%%%%%%%%%%%%%%%%%%%%%%%%%%%%%%%%%%%%%%%%%%%%%%%%%%%%%
\subsection{The Maxwell fields in the Page limit}
 \label{ss21I16.1}

	In this section we analyse the regularity of	 the one-form (\ref{15III15.7}) after passage to the limit $\epsilon \to 0$.
The coordinate transformations (\ref{10I16.1})-(\ref{10I16.3}) yield  the following form for the p-contribution of (\ref{15III15.7}) in $(\eta,\chi,\theta,\barvarphi )$ coordinates:
\bena
A^{(p)}
 &  := &  \frac{p \, \cos(\theta)}{\Sigma} \sigma_1
\nn
\\
   & =  &
   \frac{p \, \cos(\theta)}{\Xi\Sigma } \left(  a \omega_0  \left(\frac{- 2 r_0   \cos (\chi ) + O(\epsilon) }{r_0^2-a^2}  \right) \,\dnotdelta \eta  -(r^2-a^2)  \dnotdelta \barvarphi  \right)
\,.
\label{10I16.4}
\eena
Taking the Page limit, i.e. $\epsilon \to 0 $, of (\ref{10I16.4}) gives
\bena
A^{(p)} &  = &
   \frac{p \, \cos(\theta)}{\Xi\Sigma_{r_0} } \left(    - \frac{ 2  a  r_0 \omega_0   \cos (\chi )  }{r_0^2-a^2}   \,\dnotdelta \eta  -(r_0^2-a^2)  \dnotdelta \barvarphi  \right)
\,,
\nn
\eena
where
$$
 \Sigma_{r_0}=r_0^2-a^2 \cos^2 (\theta)
  \,.
$$
%.
Near $ \chi=0$
we use the $2\pi$-periodic coordinate $\phinotone $, as introduced
in analysis of the regularity of the metric, with corresponding coordinate differential
(\ref{5VIII15.11}). This gives
\bean
\nn
A^{(p)}
\nn
&   =&     \frac{p \, \cos(\theta)}{\Xi\Sigma_{r_0} } \left(   - \frac{ 2  a  r_0 \omega_0    (\cos(\chi)-1) }{r_0^2-a^2}   \,\dnotdelta \eta    -  \alphanotone  (r_0^2-a^2)  \dnotdelta \phinotone    \right)
\\
\nn
&  = &
  \underbrace{
   \frac{p \, \cos(\theta)}{\Xi } \left(  - \frac{ 2  a  r_0 \omega_0       (\cos(\chi)-1) }{\Sigma_{r_0}(r_0^2-a^2)}   \,\dnotdelta \eta + \frac{\alphanotone  a^2 \sin^2 (\theta)}{r_0^2 -a^2 \cos^2 (\theta)}  \dnotdelta \phinotone    \right)
   }_{\mathrm{smooth\ for\ } \chi<\pi}
\\
&   &
 -\frac{\alphanotone  p \cos (\theta)}{\Xi} \dnotdelta \phinotone
\,.
\eeal{21I16.11}
As in Section  \ref{ss1XII15.1}, the last term is not smooth but the resulting Maxwell field is. We also note that  the alternative potential
\bel{22I16.21}
 A^{(p)}  +\frac{\alphanotone  p  }{\Xi} \dnotdelta \phinotone =
   A^{(p)}  +\frac{  p  }{\Xi}    \bigg(\dnotdelta \barvarphi  -  \frac{2 a  r_0 \omega_0 }{  \left(r_0^2-a^2\right)^2 }   \dnotdelta\eta
  \bigg)
\ee
is smooth for $\chi<\pi$ and $\theta<\pi$, while
\bel{22I16.22}
 A^{(p)} -\frac{\alphanotone  p  }{\Xi} \dnotdelta \phinotone =
   A^{(p)}  -\frac{  p  }{\Xi}    \bigg(\dnotdelta \barvarphi  -  \frac{2 a  r_0 \omega_0 }{  \left(r_0^2-a^2\right)^2 }   \dnotdelta\eta
  \bigg)
\ee
is smooth for $\chi<\pi$ and $\theta>0$.

 \medskip

A similar analysis applies near $\chi=\pi$, and shows that the potential
\bel{22I16.23}
 A^{(p)}  +\frac{\alphanotone  p  }{\Xi} \dnotdelta \hat \phinotone =
   A^{(p)}  +\frac{  p  }{\Xi}    \bigg(\dnotdelta \barvarphi  +  \frac{2 a  r_0 \omega_0 }{  \left(r_0^2-a^2\right)^2 }   \dnotdelta\eta
  \bigg)
\ee
is smooth for $\chi>0$ and $\theta<\pi$, while
\bel{22I16.24}
 A^{(p)} -\frac{\alphanotone  p  }{\Xi} \dnotdelta \hat \phinotone =
   A^{(p)}  -\frac{  p  }{\Xi}    \bigg(\dnotdelta \barvarphi
     +
     \frac{2 a  r_0 \omega_0 }{  \left(r_0^2-a^2\right)^2 }   \dnotdelta\eta
  \bigg)
\ee
is smooth for $\chi>0$ and $\theta>0$.

 \medskip

The coordinate transformations (\ref{10I16.1})-(\ref{10I16.3}) yield  the following form for the $e$-contribution of (\ref{15III15.7}) in $(\eta,\chi,\theta,\barvarphi )$ coordinates:
\beaa
 \nn
A^{(e)} & :=  &
   \frac{e\,r}{\Sigma} \sigma_2
\\
 \nn
   &  = &
    \frac{e\,r}{\Sigma\Xi}
    \left(-   \frac{\omega_0 }{\epsilon} \frac{r_0^2-a^2 \cos^2(\theta) }{r_0^2-a^2} \dnotdelta \eta + a  \sin^2(\theta) \,   \dnotdelta \barvarphi
     \right)
\\
   &  = &
   \underbrace{ \frac{e }{ \Xi}
    \bigg(- \frac{ \omega_0  r_0}{ \epsilon (r_0^2-a^2)} \dnotdelta \eta }_{\textrm{closed}}
    +     \frac{ \omega_0   \cos \chi }{ (r_0^2-a^2)} \left(1 - 2\frac{r_0^2}\Sigma
     + O(\epsilon)
    \right)\dnotdelta \eta  +  \frac{a \, r}{\Sigma}  \sin^2(\theta)  \,   \dnotdelta \barvarphi
     \bigg)
      \,.
\eeaa
The closed part has no limit as $\epsilon$ goes to zero but can be discarded without affecting the Maxwell field. Keeping the same symbol  $A^{(e)}$ for the four-potential  obtained after removing the singular term and taking  the limit $\epsilon\to0$, we find
\bena
 \nn
A^{(e)} &  =  &
    \frac{e  }{ \Xi } \left[  \frac{\omega_0  \cos (\chi)}{(r_0^2-a^2)}\left( 1 -\frac{2 r_0^2 }{\Sigma_{r_0} }  \right)  \dnotdelta  \eta  +  \frac{a \, r_0}{\Sigma_{r_0}}  \sin^2(\theta) \,   \dnotdelta \barvarphi   \right]
     \,.
\eena
Near $ \chi=0$ we use the $2\pi$-periodic coordinate $\phinotone $, as introduced
in the analysis of the regularity of the metric, with corresponding coordinate differential
(\ref{5VIII15.11}). This yields
\bena
 \nn
A^{(e)}
     &  = &
       \frac{e  }{ \Xi } \bigg[  \frac{\omega_0  \cos (\chi)}{(r_0^2-a^2)}\left( 1 -\frac{2    r_0^2 }{\Sigma_{r_0} }  \right)  \dnotdelta  \eta
        + \underbrace{ \frac{a \, r_0}{\Sigma_{r_0}}  \sin^2(\theta) \,   \bigg( \alphanotone   \dnotdelta  \phinotone}_{\mathrm{smooth}}
         - \frac{2 a  r_0 \omega_0 }{  \left(r_0^2-a^2\right)^2 }   \dnotdelta\eta
  \bigg) \bigg]
\,.
\nn
\eena
Similarly to (\ref{1XII15.11})  the \textit{non-manifestly-smooth} part can be rewritten as:
\bena
\lefteqn{
  \frac{e \, \omega_0  }{\Xi \, (r_0^2-a^2)}
  \bigg[\left( 1 -\frac{2    r_0^2 }{\Sigma_{r_0} }  \right)   \cos (\chi)
   -\frac{2 a^2 r_0^2 \sin^2(\theta)}{\Sigma_{r_0}(r_0^2-a^2)}
  \bigg]\dnotdelta  \eta =
   }
   &&
    \nn
\\
     &  &  \underbrace{  \frac{e \, \omega_0  }{\Xi \, (r_0^2-a^2)}\left( 1 -\frac{2    r_0^2 }{\Sigma_{r_0} } \right)  ( \cos (\chi) -1) \, \dnotdelta  \eta }_{\text{\rm smooth for $\chi<\pi$}}
     -  \underbrace{  \frac{e \omega_0
   \left(a^2+r_0^2\right)}
   {\Xi
   \left(a^2-r_0^2\right)^
   2}  \dnotdelta  \eta  }_{\mathrm{closed}}
      \,,
\eena
which implies smoothness of the Maxwell field for $\chi<\pi$. We also see that the four-potential
\bel{22I16.1}
  A^{(e)}  +
  \frac{e \omega_0
   \left(a^2+r_0^2\right)}
   {\Xi
   \left(a^2-r_0^2\right)^
   2}  \dnotdelta  \eta
%      \,,
\ee
is smooth for $\chi<\pi$.

An analogous analysis near $\chi=\pi$, using the coordinate $\hat \phi$ of \eq{22I16.5}, shows that  the four-potential
\bel{22I16.6}
  A^{(e)}  -
  \frac{e \omega_0
   \left(a^2+r_0^2\right)}
   {\Xi
   \left(a^2-r_0^2\right)^
   2}  \dnotdelta  \eta
%      \,,
\ee
is smooth for $\chi>0$.

\subsection{Dirac strings}
 \label{ss22I16.1}

The results of Section~\ref{ss21I16.1} can be summarised as follows: the potential
\bel{22I16.31}
   A  +\frac{  p  }{\Xi}    \bigg(\dnotdelta \barvarphi  -  \frac{2 a  r_0 \omega_0}{  \left(r_0^2-a^2\right)^2 }   \dnotdelta\eta
  \bigg)  +
  \frac{e \omega_0
   \left(a^2+r_0^2\right)}
   {\Xi
   \left(r_0^2-a^2\right)^
   2}  \dnotdelta  \eta
\ee
is smooth for $\chi<\pi$ and $\theta<\pi$; the potential
\bel{22I16.32}
   A   -\frac{  p  }{\Xi}    \bigg(\dnotdelta \barvarphi  -  \frac{2 a  r_0 \omega_0}{  \left(r_0^2-a^2\right)^2 }   \dnotdelta\eta
  \bigg)  +
  \frac{e \omega_0
   \left(a^2+r_0^2\right)}
   {\Xi
   \left(r_0^2-a^2\right)^
   2}  \dnotdelta  \eta
\ee
is smooth for $\chi<\pi$ and $\theta>0$; the potential
\bel{22I16.33}
   A   +\frac{  p  }{\Xi}    \bigg(\dnotdelta \barvarphi  +  \frac{2 a  r_0 \omega_0}{  \left(r_0^2-a^2\right)^2 }   \dnotdelta\eta
  \bigg)
  -
  \frac{e \omega_0
   \left(a^2+r_0^2\right)}
   {\Xi
   \left(r_0^2-a^2\right)^
   2}  \dnotdelta  \eta
\ee
is smooth for $\chi>0$ and $\theta<\pi$; finally
\bel{22I16.34}
   A   -\frac{  p  }{\Xi}    \bigg(\dnotdelta \barvarphi
     +
     \frac{2 a  r_0 \omega_0}{  \left(r_0^2-a^2\right)^2 }   \dnotdelta\eta
  \bigg)
      -
  \frac{e \omega_0
   \left(a^2+r_0^2\right)}
   {\Xi
   \left(r_0^2-a^2\right)^
   2}  \dnotdelta  \eta
\ee
is smooth for $\chi>0$ and $\theta>0$.

Recall that $\bar \varphi$ and $\eta $ are $2\pi $ periodic,   and that we have (see \eq{5VIII15.1})
\bel{22I16.7}
 \frac{4 |\omega_0|       a  r_0
  }
   {
   \left(r_0^2-a^2\right)^
   2} =    n
   %
%   \,,
%   \qquad
   \in \N^*
   \,.
\ee
%.
Repeating the usual arguments as in Section~\ref{s1XII15.1}, the requirement of well defined charged Dirac fields implies existence of integers $\hat n_1$, $\hat n_2 \in \Z$ such that
\bel{22I16.6b}
   \frac{2 p q_0}
   {\hbar   \Xi}  = \hat n_1
    \,,
   \qquad
 \frac{2 \omega_0  e\left(a^2+r_0^2\right) q_0
  }
   {\hbar \Xi
   \left(r_0^2-a^2\right)^
   2} =   \hat n_2
    \,,
\ee
together with the constraint
\bel{22I16.10}
 \frac{4 |\omega_0|     p a  r_0 q_0
  }
   {\hbar \Xi
   \left(r_0^2-a^2\right)^
   2} = \frac{  n\hat n_1} 2
  \in \Z
   \,.
\ee

\bigskip

%%%%%%%%%%%%%%%%%%%%%%%%%%%%%%%%%%%%%%%%%%%%%%%%%%%%%%%%%%%%%%%%%%%%%%%%%%%%%%%%%%%%%%%%%%%%%%%

{\sc Acknowledgements:} We are grateful to M.J.~Duff for drawing our attention to \cite{DuffMadore}.  Supported in part by  the Austrian Science Fund (FWF) under project P 23719-N16 and by Narodowe Centrum Nauki under the grant DEC-2011/03/B/ST1/02625..

\bibliographystyle{amsplain}
\bibliography{../../references/hip_bib,%
../../references/reffile,%
../../references/newbiblio,%
../../references/newbiblio2,%
../../references/bibl,%
../../references/howard,%
../../references/bartnik,%
../../references/myGR,%
../../references/newbib,%
../../references/Energy,%
../../references/netbiblio}

\end{document}